# Human-Machine Collaboration for Automated Vehicles via an Intelligent Two-Phase Haptic Interface

## Authors

Chen Lv[1]*, Yutong Li[1], Yang Xing[1], Chao Huang[1], Dongpu Cao[2], Yifan Zhao[3], Yahui Liu[4]

## Affiliations

1. School of Mechanical and Aerospace Engineering, Nanyang Technological University, 50 Nanyang Ave, 639798, Singapore
2. Mechanical and Mechatronics Engineering, University of Waterloo, ON, N2L 3G1, Canada
3. School of Aerospace, Transport and Manufacturing, Cranfield University, Bedford, MK43 0AL, UK
4. School of Vehicle and Mobility, Tsinghua University, Beijing 100084, China

* Corresponding author. Email: lyuchen@ntu.edu.sg

## Abstract

Prior to realizing fully autonomous driving, human intervention will be required periodically to guarantee vehicle safety. This fact poses a new challenge in human-machine interaction, particularly during control authority transition from the automated functionality to a human driver. This paper addresses this challenge by proposing an intelligent haptic interface based on a newly developed two-phase human-machine interaction model. The intelligent haptic torque is applied on the steering wheel and switches its functionality between predictive guidance and haptic assistance according to the varying state and control ability of human drivers, helping drivers gradually resume manual control during takeover. The developed approach is validated by conducting vehicle experiments with 26 human participants. The results suggest that the proposed method can effectively enhance the driving state recovery and control performance of human drivers during takeover compared with an existing approach, further improving the safety and smoothness of the human-machine interaction in automated vehicles.

## MAIN TEXT

The development of automated roadway vehicles has generated increasing attention from both academia and industry in recent years. However, the development of highly automated vehicle or semi-autonomous vehicle, where the driving task is exchanged periodically between human drivers and automated vehicle technologies, can be expected to precede a transitioning to fully autonomous vehicles (*1*–*3*). The transition between human-driving and automated-driving modes represents a particular risk because human drivers may be preoccupied with a non-driving activity (NDA), and some time may be required for humans to recover a suitable level of driving performance required for assuming control safely (*4*–*6*). As such, guaranteeing safe, smooth, and swift control authority transitions between a human driver and the automated functionality of the vehicle is one of the critical issues of this technology (*7*–*9*). This challenge requires new cross-disciplinary theory, analysis, and design approaches related to human-machine collaboration.

To address this issue, many studies have been conducted from a human factors perspective to investigate the key factors which have an influence on the takeover performance. These key factors mainly include the required takeover time, the modality of the takeover request (TOR) signal, secondary task engagement, driver states, and driving conditions (*10*–*12*). The effect of the takeover time upon driver reactions and control performance was investigated (*13*). The results showed that with shorter TOR time, the participants reacted faster, but the takeover control performance became worse. The increase of the takeover time was found to be heavily related to the level of the cognitive workload occupied by the secondary tasks (*14, 15*). Different modalities of the TOR signal were investigated in (*16*), and the results showed that users' preferences for TOR modalities in highly automated vehicles depended on the urgency of the driving situation. The impact of traffic densities upon the takeover process was explored, and the results indicated that the high density of the traffic flow would have a negative impact on both takeover time and post takeover performance (*17*). Besides, driver's readiness and takeover ability were also explored from the angle of modelling and estimation (*10, 18–20*).

In addition to the above human factor studies, advanced control methods have also been adopted to solve human-machine interaction issues for automated vehicles. In some studies, the takeover strategy of instant control transfer from automation to driver was adopted even the human may not be ready for the required driving task (*13, 21*), while a period of shared control was suggested as a promising solution to further enhancing vehicle safety and comfort during handover (*22*). Leveraging on a novel definition of cooperative state between automation and driver, a smooth control authority transfer from automated driving to manual driving with haptic shared control was developed (*23*). The proposed strategy of authority transfer was realized by tuning a design gain which was correlated to the driver's steering torque. The effects of different types of haptic steering torque upon driving performance were also studied (*12*, *24*). The results indicated that a continuous haptic steering torque can improve the path following performance of drivers during standard steering maneuvers. From literatures, there are many possible ways to handle driver-automation collaboration, including handover, for automated vehicles by utilizing haptic shared control. However, some of these methods are not associated with a model of the human driver. Taking a prediction of driver's action into consideration may further benefit the performance of human-machine cooperation. The intention of the driver was taken into account within the modeling and included in the objective function to minimize controller intervention during driver-automation shared control (*25*). With the similar idea, some frameworks were also proposed and implemented in the design of haptic shared control and advanced driver assistance system by integrating a dynamic model of driver (*26–28*), but the authorities allocated to human and automation were fixed. Extended from the above concepts, a new framework was developed based on a game-theoretical model of human-machine interaction for dynamic role distribution (*19, 29*). Under this framework, different concepts of control transitions were developed for takeover of automated vehicles and compared via human-in-the-loop experiments (*30*). The test results indicated that the shared control based methods were advantageous over those ones with immediate shutdown of automation. However, since the human behavior was described by an optimal controller, the imperfect driver behavior during the takeover process could hardly be captured. In addition, the dynamic adaptation of handover parameter, as well as real-vehicle implementation and validation have not been fully addressed yet.

Although many achievements have been made in the past, challenges in human-machine interaction of automated vehicles, particularly for the automation-to-driver takeover process, still remain, and these require the design of novel human-machine collaboration systems. During takeover, humans may need some time to recover from pre-occupied secondary tasks to a suitable level of required driving performance, while their mental and physical states and readiness are varying. Clearly, some types of guidance and assistance to human drivers are needed to ensure the safe, smooth, and swift completion of the handover. Thus, to further advance the approach for takeover control of automated vehicles, the present work develops a human-machine collaboration approach that provides necessary guidance and assistance to human by using an intelligent two-phase haptic interface, which is expected to adapt to varying state and control ability of drivers during takeover.

**Fig. 1. Architecture of the human-machine collaboration system with the intelligent haptic interface for automated driving.**

As shown in Fig. 1, during takeover transition, the proposed human-machine collaboration system modulates the automation system's control effort according to the measured states and control action of a driver, and applies an intelligent haptic steering feedback to the driver to ensure that the takeover transition is completed in a safe and smooth manner. The detailed control block diagram of the system is shown in Supplementary Fig. 1. Rather than a simple vibration, the intelligent haptic interface proposed is a torque. As shown in Fig. 2, the functionality of the intelligent haptic torque is divided into two phases, namely phase 1: haptic guidance, and phase 2: haptic assistance. The haptic guidance torque is provided when driver's control capability is not ideal and expected to guide the human driver to properly operate the steering wheel, actively helping the recovery of driver's situation awareness and driving ability. When driver's control capability recovers to a high level, the functionality of the haptic interface switches from guidance to assistance, only providing slight corrections and assisting to smooth vehicle trajectory until the takeover being completed. The detailed experimental results and methodology adopted are reported as follows.

**Fig. 2. Schematic diagram of the takeover process under the intelligent two-phase haptic interface.**

## Results

The feasibility and effectiveness of the proposed intelligent haptic takeover control was investigated by conducting experiments in an automated vehicle (Figs. 3a and b) involving 26 participants engaged in two tasks, namely automation-to-driver takeover control under a single-lane normal steering condition (Task A; Fig. 3c), and automation-to-driver takeover control under a single lane change maneuver (Task B; Fig. 3d). For comparison, experiments were completed by each participant for each task under two different haptic takeover methods. One method was the proposed one, which was using an intelligent two-phase haptic feedback (Fig. 2), and the other one was the baseline approach, which was with a fade out of the autopilot steering torque (Supplementary Fig. 2). The order of the experiments for each participant was randomized.

**Fig. 3. Experimental set-up.**

During the experiments, the allowed driver control authority $\alpha^{ref}$, the degree of driver intervention or control performance $\alpha$, the haptic guidance torque $T_{hpt}$, the torque applied to the steering wheel by the driver $T_H$, and the yaw rate $\dot{\psi}$ of the vehicle, were recorded for each participant over time (refer to the Section of Methods for detailed definitions of these variables). Example data of the automation-to-driver takeover process for a representative participant while conducting Task A are shown in Fig. 4. Experimental results for a representative participant while conducting Task B were illustrated in Supplementary Fig. 3. Further statistical analysis of the measured data was conducted for all 26 participants under the designated tasks based on the takeover time, driver steering torque, and yaw rate of the vehicle (Supplementary Tables 1 and 2). The values are presented in the form of the mean ± the standard deviation (SD). The time span used to calculate the mean values of the measured signals was each participant's individual takeover time.

| Fig. 4. Example data for a representative participant while conducting Task A with baseline and proposed methods. |
|---|

### Takeover time

The takeover time was recorded and assessed for each participant under baseline and proposed methods (Supplementary Tables 1 and 2). According to the results shown in Figs. 5a and 6a., for the test with baseline method, the mean value of the takeover time values of all participants was 8.0 ± 0.6 s for Task A and was 7.9 ± 0.8 s for Task B. While for the test with the proposed method, their mean was 4.4 ± 0.2 s for Task A and was 4.4 ± 0.3 s for Task B. The statistical significance of differences in mean values of the takeover time under the two approaches was further analyzed via paired t-test. Based on the results, the reduction in takeover time under the proposed method can be regarded as statistically significant ($p$<0.01).

| Fig. 5. Box plots of the experimental results of Task A. | Fig. 6. Box plots of the experimental results of Task B. |
|---|---|

### Driver steering torque

The steering torque $T_H$ applied by the driver was recorded and assessed for each participant during the takeover process under baseline and proposed methods (Supplementary Tables 1 and 2). As shown in Figs. 5b and 6b, for the baseline group, the mean value of the average $T_H$ of all participants was 0.9 ± 0.21 N·m for Task A and was 0.33 ± 0.1 N·m for Task B. While under the proposed method, the mean of the average values of $T_H$ of all participants was 0.8 ± 0.07 N·m for Task A and was 0.28 ± 0.04 N·m for Task B. The statistical significance of difference in SD values of $T_H$ under the two approaches was analyzed using paired t-test. For both Tasks A and B, the reductions in the SD values of $T_H$ under the proposed method were considered as statistically significant ($p$<0.01), compared to those ones obtained in the baseline group.

### Steering wheel angle

The steering wheel angle was recorded and assessed for each participant (Supplementary Tables 1 and 2). As shown in the box plots of Figs. 5c and 6c, for the baseline test, the mean of the average values of $\theta_{sw}$ across all participants was 13.9 ± 1.7 degrees for Task A and was 6.44 ± 1.6 degrees for Task B. While under the proposed method, their mean value was 14.4 ± 0.7 degrees for Task A and was 6.39 ± 0.6 degrees for Task B. Significant reductions in SD values of $\theta_{sw}$ under the proposed method was identified via paired t-test ($p$<0.01), compared to the results obtained for the baseline group.

### Yaw rate of the vehicle

The yaw rate of the testing vehicle was recorded and assessed for each participant during the takeover process (Supplementary Tables 1 and 2). As shown in Figs. 5d and 6d, under the baseline approach, the mean of the average values of the yaw rate across all participants was 2.3 ± 0.5 deg/s for task A and was 0.8 ± 0.37 deg/s for Task B. While under the proposed method, the mean of the average values of the yaw rate across all participants was 2.1 ± 0.17 deg/s for Task A and was 0.7 ± 0.07 deg/s for Task B. The statistical significance of reduction in SD values of the yaw rate under the proposed method was also observed via paired t-test ($p$<0.01), compared to the results for the baseline group.

## Discussion

A comparison of results shown in Fig. 4a indicates that the vehicle trajectory was more consistent under the developed intelligent haptic interface compared to that obtained with baseline approach, although the representative participant successfully completed Task A while staying within the lane under both conditions. Figs. 4b to f demonstrate that the baseline strategy resulted in the driver making many oscillations in the steering torque and angle, and a relatively long time was required to complete the task. Meanwhile, under the proposed strategy as shown in Figs. 4g and h, the driver applied the required steering torque smoothly under the guidance and assistance provided by the proposed intelligent haptic interface. This resulted in the driver implementing full manual control within a shorter period of time with little fluctuation in the steering torque and angle, as shown in Figs. 4i to k. Similar results were obtained for the same representative participant while conducting Task B (Supplementary Fig. 3). A statistical analysis of the data like that shown in Fig. 4 was conducted for all 26 participants under the designated tasks based on the takeover time, driver steering torque, steering wheel angle, and yaw rate of the vehicle (Supplementary Tables 1 and 2).

Takeover time is an important metric for assessing a driver's takeover ability and takeover performance. A short takeover time could be essential for mitigating the risks associated with particular driving scenarios during the automation-to-driver takeover process in an automated vehicle. In literatures, the takeover time is usually considered as the time span between the TOR and the maneuver start as a reaction to the system limit (*10*). The threshold values that determine the start of the maneuver were adopted as 2° steering wheel angle change and 10% brake pedal actuation (*13*). However, the start of the maneuver may be not a reasonable condition to indicate the completion of takeover, since at the initial stage the driver could lack situation awareness and therefore possess low qualification for safely driving. Instead, we maintain that not only starting a control action, but also the human driver achieving good control performance during control transition, should be considered as the completion of the takeover. Thus, in this study, we define the takeover time as the elapsed time between the TOR and the first stabilization of the steering operation (refer to the Section of Statistical Analysis for detailed definition). According to the results shown in Figs. 5a and 6a, the intelligent two-phase haptic interface generated by the developed human-machine collaboration system reduced the mean takeover time for Tasks A and B by 51.25% and 44.30%, respectively, compared with the baseline approach.

Takeover control performance is usually related to the difficulty of the required driving task, the cognitive and physical states of the driver, and the skill and experience of the driver. The mean values of the driver steering torque $T_H$ and the steering wheel angle $\theta_{sw}$ reflect the required effort of the driving task, while the SD is indicative of the consistency of the control performance of each individual driver during the takeover process. It should be noted that the mean values of $T_H$ and $\theta_{sw}$ for the baseline and proposed methods in each experimental task were similar according to the results shown in Figs. 5b, 5c, 6b and 6c. This was because the required action, i.e. tracking lane centerline in Task A, or conducting lane change in Task B, was the same for both the test groups with baseline and proposed methods. However, the identified reductions in SD values of $T_H$ and $\theta_{sw}$ under the proposed method suggest that the haptic guidance and assistance provided by the developed system mitigates the impacts of variations in the states and behaviors of the individual drivers on their takeover control performance, enhancing the consistency.

The yaw rate $\dot{\psi}$ is indicative of vehicle maneuverability during the takeover process. The SD values of the yaw rate under the proposed method were found to be significantly reduced compared to those obtained for the baseline group. The relatively high variability over the yaw rate value of each participant in the baseline group also demonstrates the effects of variations in individual driver characteristics and behaviors. The comparison of the results obtained for the two groups demonstrates that the takeover control performances of the drivers were much more consistent under the provided intelligent haptic interface, and thus ensured the maneuverability of the vehicle during the takeover process.

In addition, survey of all participants was conducted after their test runs. The survey results shown in Supplementary Fig. 4 revealed that the proposed intelligent haptic interface led to a slightly different feeling during takeover transitions but was still regarded as pleasant.

The above experimental results suggest that the proposed intelligent haptic interface can help speed up the driver's driving state recovery and improve the manual control capability during the takeover process. Besides, the high-level control framework, methodology employed, as well as models developed in this work could be expanded to a wide range of human-machine interaction applications.

The design of a human-machine collaboration system for automated vehicles is a system engineering task that requires the development and cooperation of a number of different areas such as human factors, control engineering, signal processing, ethics, and laws. In the present study, the cognitive and physical states of a driver were considered as discrete levels rather than as more continuous levels. As such, the coarseness of this discretization may limit the smoothness of driver state assessment, as well as the smoothness of the allowed control authority allocation. To further improve the human-machine collaboration quality, quantitative evaluation of driver states with parameter sensitivity should be investigated in the future. In addition, limiting driver state assessment to include only attention and neuromuscular states may not be a sufficiently complete assessment of the cognitive and physical statuses of drivers. Therefore, additional signals reflecting the psychological and physiological states of drivers would be included in future studies. We also employed a fixed modality, intensity, and frequency for the TOR signal in the conducted experiments. However, this may restrict the possible range of reaction sensitivities available to drivers engaged in different NDAs during the takeover transition. Therefore, adopting a multi-modal TOR signal that can adapt to the different non-driving activities of drivers should be explored. Besides, the experiments conducted in this work only focused on normal driving conditions, and emergency takeover under critical situations were not considered.

## Methods

### Experimental design

The experimental sport utility vehicle shown in Fig. 3a was modified and used as the testing platform for a range of experiments in automated driving. Technical details with specifications of the testing vehicles are reported in Supplementary Note 1 and Supplementary Table 3.

Two tasks were assigned to the 26 participants (described in detail below), including Task A: automation-to-driver takeover control under a single-lane normal steering condition, and Task B: automation-to-driver takeover control under a single lane-change maneuver. All experiments were conducted in a certified testing area involving three vehicle lanes that were each a uniform width of 3.5 m. Each participant was asked to conduct both Task A and Task B with both takeover control methods. One strategy was the proposed intelligent haptic feedback (Fig. 2), while the other one, i.e. the baseline method, was using a fade out of the autopilot torque with a fixed slope of 2.5 N·m/s (Supplementary Fig. 2). For each task, each participant was firstly asked to naturally drive the testing vehicle for five minutes to get used to the car, and then the participant was required to complete one run for each of the takeover methods. The required tasks as well as the adopted takeover strategies within the experiments for each participant were randomized in order to avoid learning effects.

The goal of both tasks was for the driver to resume steering control after perceiving a multi-modal TOR signal (described in detail below), while ensuring that the designated driving maneuver was accomplished. Before the experiments with the proposed method, the participants were informed that the steering system would provide haptic feedback during the takeover transition. All experiments began with the vehicle stationary on the three-lane roadway. Then, the experimental vehicle was driven in the self-driving mode by the automation system by tracking the centerline of the middle lane of the three-lane roadway, and the vehicle accelerated to the target cruising speed of 10 m/s. Throughout this period, the human driver was instructed to disregard the roadway and read news on a mobile phone. This designed NDA was a cognitively, visually and physically demanding one. The experimenter sat on the rear passenger seat, and activated the automatic cruise and lane keeping functions, thereby enabling autonomous driving in both longitudinal and lateral directions. For Task A, the roadway for testing had a certain curve with the estimated radius of 190 m. And the TOR signal was triggered automatically by vehicle localization signal at a pre-defined position. For Task B, the left-turn signal was first engaged at a designated position of a straight section of the roadway. This was a command for the automated driving system to change lanes from the current middle lane to the left lane. And the TOR signal was automatically triggered after 0.5s of the left-turn signaling. After perceiving the TOR, the human driver was asked to put the mobile phone down immediately and turn their attention to the driving task in preparation for resuming control of the vehicle. Once the driver was identified to have their hands firmly on the steering wheel, the transition of the control authority was triggered. Automatic cruising was engaged throughout all experiments. Thus, each participant was only required to focus on steering control during the takeover action. To ensure the consistency of the experiments, the drivers were informed in advance to initiate the takeover action as soon as possible once perceiving the TOR signal. After completion of the maneuver, the driver was asked to depress the brake pedal and bring the vehicle to a complete stop. The subsystems, methods, and algorithms that make up the above tasks are described in detail as follows.

In addition, each participant was required to complete a questionnaire after their own test runs to gather his or her personal opinion. The evaluation of the subjects was captured by two questions. They were asked about whether they have noticed a difference between the proposed haptic takeover and the baseline approach, as well as the steering feeling of the proposed one. Both categories were rated on a scale from one (no difference/very unpleasant) to five (very different/very pleasant).

## Takeover Request Signaling

The multi-modal TOR signal was comprised of visual and auditory components that were activated simultaneously. The visual request signal was the text “Please take over!” shown on the dashboard (Supplementary Fig. 5) until the end of the handover transition. The auditory signal was a 70 dB beep emitted at a frequency of 5 Hz lasting for 0.5 s.

## Assessment of Driver States and Control Ability

In the proposed human-machine collaboration system, the control ability of the human driver is associated with cognitive and physical states that are assessed online in real time. To simplify the implementation in this study, the focus of driver attention and the muscle state in the upper limbs were adopted as the indicators of cognitive and physical states, respectively.

For cognitive state assessment, the onboard driver monitoring system (Supplementary Note 1) detects the driver’s body pose (i.e., driving or non-driving activity), the gaze movement, the blink frequency, and etc., and thereby comprehensively assesses the current level (high or low) of the driver’s attention to the driving activity. Here, driver attention is deemed low when the driver’s current behavior reflects a non-driving or distracted state, such as when the driver is looking down at a mobile phone. In contrast, driver attention is deemed high when the driver’s current behavior reflects a normal driving pattern, such as, when reacting to a TOR, the driver puts the mobile phone down and transitions to the driving task, checking mirror and surrounding vehicles.

Muscle state can be represented by the neuromuscular dynamics of the driver’s arms during steering operations. During the takeover process, a relatively large steering torque may be required to maneuver the vehicle, which would necessitate a relatively large degree of muscle state compared to the more complete muscle relaxation of the non-driving state. To quantitatively assess the degree of muscle state, the neuromuscular dynamics of the driver's arms were subjected to characterization and parameterization. To this end, the coupled system of the driver and steering system was abstracted into the following model (*32-34*):

$$G_s(s) = \frac{\theta_{sw}}{T_H} = \frac{1}{(J_{dr} + J_{st})s^2 + (B_{dr} + B_{st})s + (K_{st} + K_{dr})}, \tag{1}$$

where $\theta_{sw}$ is the angular position of the steering wheel, $T_H$ is the torque applied to the steering wheel by the driver, $J_{dr}$, $B_{dr}$, and $K_{dr}$ are the inertia, viscous damping coefficient, and the stiffness coefficient of the driver, respectively, and $J_{st}$, $B_{st}$, and $K_{st}$ are the inertia, viscous damping and stiffness coefficients of the steering system, respectively. Existing studies have reported that the value of $K$ is highly correlated with muscle activity during driving, where increasing $K$ reflects increasing muscle activity (*32-34*). Thus, $K$ is selected as the key indicator of muscle state. The actual value of $K$ can be estimated online by using the reported methods in (*33-35*). Thus, the level of muscle state (i.e. a driver's physical state) can be considered high when $K$ exceeds a predefined threshold $K_1$, which was set as 2.5 N·m/rad in the experiments. Otherwise, the driver's physical state is considered to be low, which is indicative of being unqualified for engaging in manual driving.

In the present work, the control ability of a human driver is comprehensively evaluated as low, medium, and high based on the measured levels of the driver's cognitive and physical states according to the scheme illustrated Supplementary Fig. 6.

## Driver Control Authority and Performance

To ensure a safe and smooth transition of control from the autonomous driving mode to the manual driving mode, the maximum allowed control authority $\alpha^{ref}$ ($0 \le \alpha^{ref} \le 100\%$) of a human driver is gradually increased and released to the driver, setting the upper bound for the degree of driver's intervention. In the baseline method, the control authority will be completely transferred to the driver once after the driver intervenes in the control. While in the proposed approach, the total driving authority will be gradually allocated to the driver and completely transferred to the driver only after he or she exhibits an ability that fully qualifies for manual driving. Therefore, in this work, based on drivers' different cognitive and physical states, we divide their control ability into three discrete levels, namely low, medium, and high, respectively. Corresponding to the above three levels of driver control ability, the allowed driver control authority $\alpha^{ref}$ is further set as 30%, 60%, and 90% or 100%. The detailed mechanism for determining $\alpha^{ref}$ based on driver's states is illustrated in Supplementary Fig. 6.

In addition, another important state variable $\alpha$ ($0 \le \alpha \le 100\%$) indicating the driver's control performance and the degree of intervention (or the actual control authority taken by the driver) during takeover is defined here as:

$$\alpha = \min(\alpha^{ref}, \frac{T_H}{T_{ref}}), \tag{2}$$

where $T_{ref}$ is the optimal system input torque that is sufficient to ensure that the vehicle tracks the lane centerline. In this work, we assume that the automation system is still working and able to calculate $T_{ref}$ within the allowed time budget for takeover. The detailed method and parameters used to compute $T_{ref}$ are reported in Supplementary Note 2, and Supplementary Tables 3 and 4.

In the baseline approach, $\alpha$ can be seen as an indicator of the driver's control performance. While in the proposed method, $\alpha$ indicates the actual degree of driver's intervention. In this work, three threshold parameters of $\alpha$ (shown in Fig. 2a) are designed as $\alpha_1$=30%, $\alpha_2$=60%, and $\alpha_3$=90%. After exceeding the pre-defined threshold $\alpha_3$, if and only if $\alpha$ steadily holds within the interval between $\alpha_3$ and 100% for a period of time (set as 1.5s in this work), the vehicle can be considered as stabilized and fully controlled by the human driver. Then the control authority will be entirely transferred to the human driver, and the takeover can be seen as completed.

It should be noted that the presented concept of $\alpha$ for assessing the degree of driver's intervention will only work if the required steering torque is significantly unequal to zero. However, in a takeover scenario that happens on a straight road, $\alpha$ may hardly be calculated. In that case, indicators that could effectively reflect the driver's takeover performance need to be further explored. Some simple but reliable determination criteria could be applied. For example, if the driver puts his hands on the steering wheel, and the steering wheel angle holds within a small interval that is close to the neutral position for a period of time, then the takeover could be seen as completed.

## Modeling of Human-Machine Collaboration Process

The predictive haptic guidance and assistance controller is designed on the basis of a human-machine interaction model. The model must sufficiently describe the interactive behaviors and the shifting roles between human and automation control under haptic guidance and assistance. In this study, we model driver-automation collaboration as a two-phase process that includes automation dominance and human dominance.

*1) Automation dominance.* The driver control ability is low in the early stage of takeover, resulting in a low value of $\alpha^{ref}$. Therefore, automated control should dominate during this stage, and only a small portion of the control authority should be allocated to the human driver.

*2) Human dominance.* The driver control ability recovers to a relatively high level in the later stage of takeover, and the value of $\alpha^{ref}$ is accordingly increased. Therefore, human driver control should dominate during this stage, with the contribution of automation control being decreasing accordingly.

Here, automation dominance is assumed if $\alpha$ is less than the designed threshold $\alpha_2$, which is set as 60% in the present study. Otherwise, human dominance is assumed.

As presented in the high-level control framework of the system (Supplementary Fig. 1), the input contributed by automation system $T_A$ always compensates for the summation of driver's actual torque and the haptic torque $T_{hpt}$ during the takeover process, occupying the remaining part of the optimal control input $T_{ref}$. It will be directly applied to the downstream vehicle plant rather than the steering wheel to ensure that the vehicle is able to track the expected lane centerline. Thus, the automation's torque $T_A$ can be expressed as

$$T_A = T_{ref} - T_H - T_{hpt}\,. \quad (3)$$

The detailed method of implementing $T_A$ and $T_{hpt}$ in the experimental car is reported in Supplementary Note 1.

The overall input $u$ to the physical plant of the vehicle is expected to be consistent with the required optimal input $T_{ref}$, and it can be calculated as

$$u = T_H + T_A + T_{hpt}\,, \quad (4)$$

where $T_{hpt}$ is the haptic feedback torque. The detailed control algorithm for $T_{hpt}$ is introduced in the following section.

## The Two-Phase Predictive Haptic Steering Torque Controller

According to the automation dominance and human dominance phases defined above, the functionality of the haptic takeover controller is classified as: Phase 1: predictive haptic guidance, and Phase 2: haptic assistance, respectively. Disengaging from the pre-occupied NDA and transitioning back to the driving task, when driver's control ability is medium and below, it is still in the phase of automation dominance. Thus, i.e. the phase 1, the haptic feedback torque should be generated based on the estimation of driver's future reaction using a human model of the automation dominance phase. In this phase, the haptic torque applied on the steering wheel is expected to guide the driver to properly control the steering wheel to the suitable position, and simultaneously help recover the situation awareness. As the state and control performance of the driver recover, phase 2 starts, i.e. the human dominance. Therefore, in phase 2, the functionality of the haptic feedback transitions from guidance to assistance. The assistive torque provided initiates only slight corrections consistent with the operations of the driver, compensating for driver's imperfect action, smoothing vehicle trajectory, and further improving driver's control performance. Thus, the intelligent two-phase haptic controller is designed as follows.

*1) Predictive haptic guidance.* An appropriate value of $T_{hpt}$ is generated and applied to the steering system for guiding the human driver to steer the vehicle in the proper direction, and for also imparting a better understanding of the required driving task. Assuming that the human driver follows the guidance provided by $T_{hpt}$, we can describe the dynamics of interaction between haptic guidance and human action as

$$\lambda T_{hpt} = \tau_H \dot{T}_H + T_H \,, \tag{5}$$

where $\tau_H$ is a time constant representing the driver's reaction time, and $\lambda$ is a gain representing the amplified influence of $T_{hpt}$ on the driver's activity. To achieve the goal of predictive haptic guidance, an optimization problem is formulated to compute $T_{hpt}$ (*36*). Here, the objective function given in equation (6a) is minimized according to the deviation between $\alpha^{ref}$ and $\alpha$, while being subject to the constraints given by equations (6b)-(6e).

$$\min_{T_{hpt,0|k}} \sum_{i=1}^{N} \left\| \alpha_{i|k} - \alpha_{i|k}^{ref} \right\|_W^2 + \sum_{i=0}^{N-1} \left\| T_{hpt,i|k} \right\|_Q^2 \tag{6a}$$

$$\text{s.t. } T_{H,i+1|k} = f(T_{H,i|k}, T_{hpt,i|k}) \tag{6b}$$

$$T_{hpt,\min} \le T_{hpt,i|k} \le T_{hpt,\max} \tag{6c}$$

$$\Delta T_{hpt,\min} \le \Delta T_{hpt,i|k} \le \Delta T_{hpt,\max} \tag{6d}$$

$$\alpha_{0|k} = \alpha_k, \;\; T_{hpt,-1|k} = T_{hpt,k-1} \tag{6e}$$

where, the dynamics model equation (6b) is obtained by discretizing equation (5) using the Euler method (*37*). The above optimization problem is solved by using MPC with a moving horizon. $N$ is defined as the prediction horizon, $W$ and $Q$ are weighting factors, and $\alpha_{i|k}$ denotes the $i^{\text{th}}$ state prediction at time step $k$ obtained by applying the optimal input $\mathbf{u}_k=\{T_{hpt,0|k}, T_{hpt,1|k}, \ldots, T_{hpt,N\text{-}1|k}\}$ to equation (6b) beginning from the measured state $T_{hpt,k-1}$ in equation (6e) at the current time step $k-1$. In addition, the constraints on $T_{hpt}$ are explicitly considered in equations (6c) and (6d). The value selections of the key parameters used in the MPC are listed in Supplementary Table 5.

*2) Haptic assistance.* The value of $T_{hpt}$ should be reduced from a guidance role to an assistance role that is consistent with the activities of the human driver, and thereby correcting driver activity via compensation. In this functionality, $T_{hpt}$ is designed as

$$T_{hpt} = \alpha^{ref} T_{ref} - T_H \,. \tag{7}$$

If the level of driver's control ability is considered as high, when the degree of driver's intervention $\alpha$ exceeds the threshold $\alpha_3$=90% (100 will be assigned to $\alpha^{ref}$, refer to Supplementary Fig. 7 for details), and $\alpha$ steadily holds between 90% and 100% for 1.5 s, then the human driver is considered fully qualified for the required driving task. At this time, the takeover control transition is deemed completed, the haptic steering torque ceases, and the automation system is entirely disengaged.

## Participants

A total of 26 participants (16 males, 10 females) in the age range of 22 to 50 (mean = 31.08, SD = 7.23) were recruited for the experiments. Each participant had a valid driving license and signed an informed consent form. The study protocol and consent form were approved by the Nanyang Technological University Institutional Review Board. All participants had no previous knowledge of the research topic and had never previously experienced haptic takeover during driving. Before experiments, the participants were informed that the steering system would provide haptic feedback during the takeover transition.

## Statistical Analysis

Statistical analysis of the experimental data was conducted under the designated tasks based on four metrics.

*1) Statistical methods.* The statistical analysis was performed in Matlab (R2017b, MathWorks) using the Statistics and Machine Learning Toolbox and in Microsoft Excel. Statistical significance was determined using paired t-tests at the $\alpha = 0.01$ threshold level throughout the paper. Central tendency was estimated using the mean.

*2) Definition of the evaluation metrics.* Four metrics were adopted to evaluate the takeover performance. The first one is takeover time. In the present work, we define the takeover time as the elapsed time between the TOR and the first stabilization of the steering operation. The first stabilization of the steering operation is defined as the point in time when $\alpha$ attains a value between 90% and 100% and sustains for 1.5 s. Here, the takeover time is a key parameter reflecting the speed with which a human driver achieves a good driving performance from being initially preoccupied with an NDA. The second metric is the driver steering torque, which is applied by the human driver on the steering hand wheel. It directly reflects the driver's action during the takeover process. The third metric is the steering wheel angle, which is the angular movement of the steering hand wheel. Since it is resulted by both the human and automation's actions, thus it can indicate the interactive behavior between the human and machine. The last metric is the yaw rate of the vehicle $\dot{\psi}$, which is the first derivative of the yaw angle of the vehicle (Supplementary Fig. 7), reflecting vehicle maneuverability.

## Data availability

All data generated or analysed during this study are included in this published article (and its Supplementary Information files).

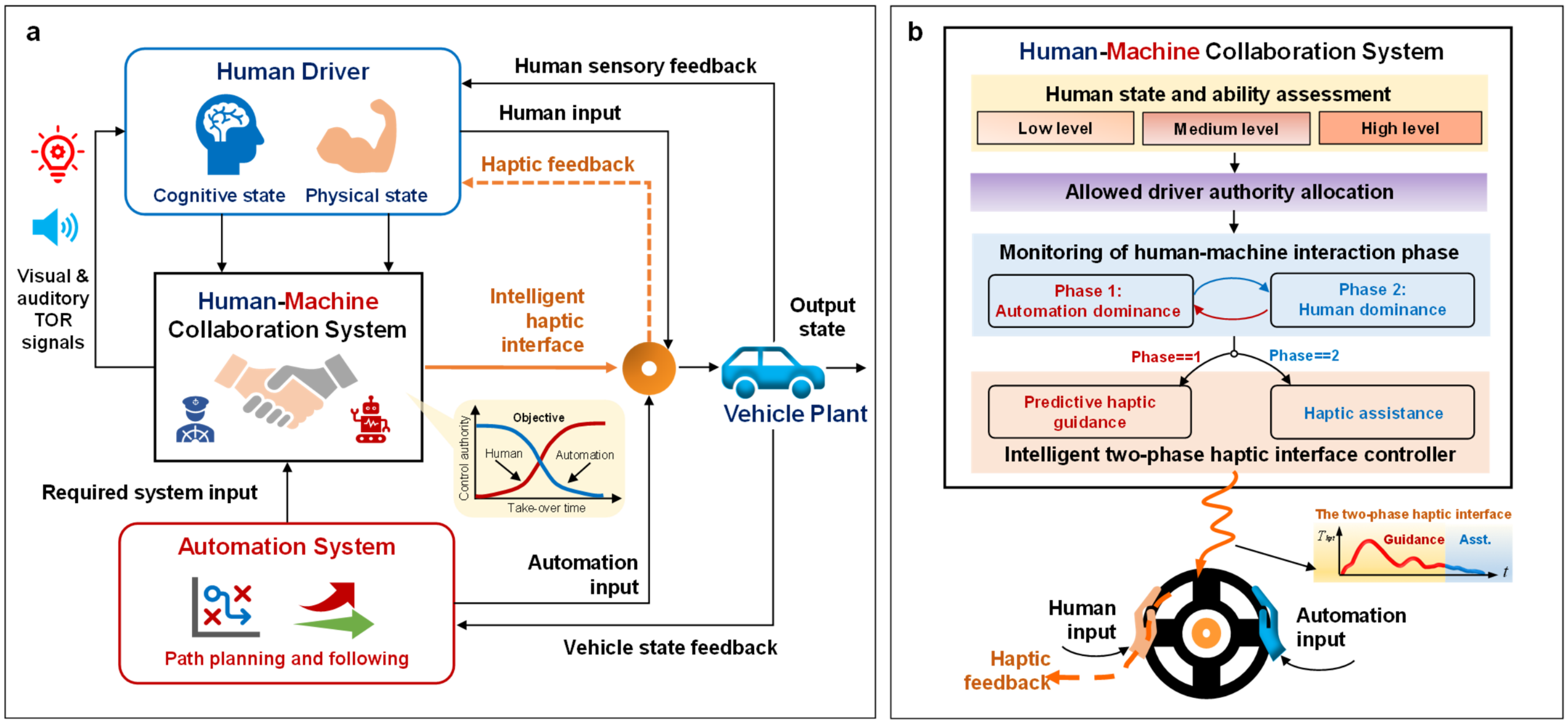

**Fig. 1. Architecture of the human-machine collaboration system with the intelligent haptic interface for automated driving. a**, After perceiving the multi-modal TOR signal, the human driver is required to take over control of the vehicle driven in the autonomous mode. During takeover transition, the proposed human-machine collaboration system modulates the automation's control efforts according to the driver's states and control action, providing an intelligent haptic steering interface to help human driver take over control in a safe and smooth manner. With the gradual increase of driver's input, automation's input decreases accordingly, and the handover process is expected to be completed gradually. **b**, The proposed system assesses the states and control ability of the driver in real time, deciding the maximum control authority (the upper bound) that could be allocated to driver. In the meantime, based on the monitored status of the human-machine interaction process, the two-phase intelligent haptic torque is applied on the steering wheel. If the automation system dominates the control (in phase 1), then a haptic guidance torque will be provided based on the prediction of driver future behavior, helping human driver apply appropriate degree of steering torque in an appropriate manner. If the driver starts to dominate the control (in phase 2), then the functionality of the haptic interface will be switched from predictive guidance to assistance.

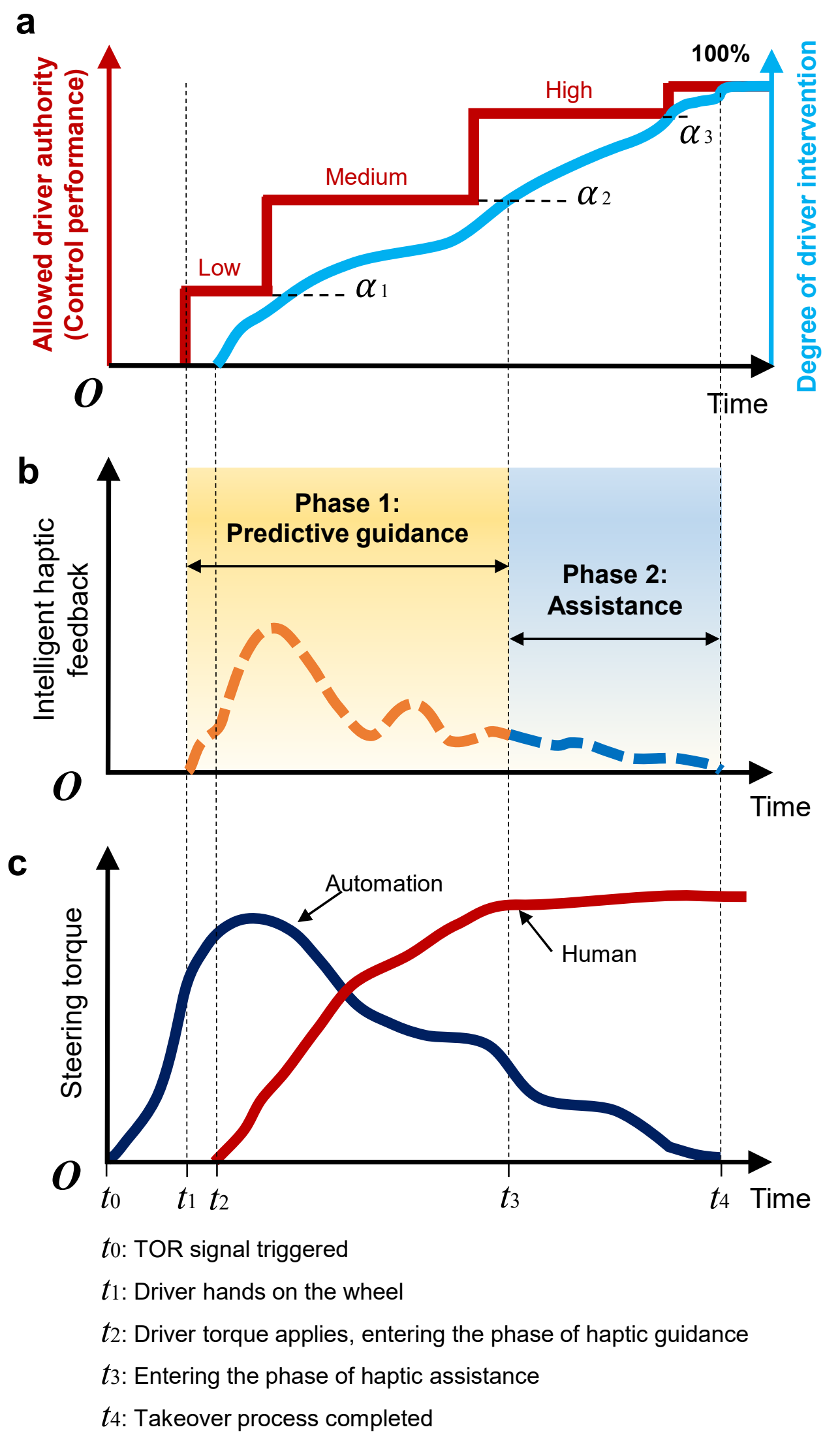

**Fig. 2. Schematic diagram of the takeover process under the intelligent two-phase haptic interface. a**, The allowed control authority gradually increases, with the degree of driver intervention and control performance being increasing under provided haptic feedback. **b**, Disengaging from the pre-occupied NDA and transitioning back to the driving task, when driver's control ability is medium and below (phase 1), the predictive haptic guidance torque is generated, guiding driver to properly steer the hand wheel to the suitable position, gradually recovering the situation awareness and the manual control ability. As the states and control performance of the driver recover (phase 2), the functionality of the haptic feedback transitions from guidance to assistance at $t_3$ and only provides slight correction consistent with the operations of the driver, compensating for driver's imperfect action and smoothing vehicle trajectory. **c**, The driver perceives the TOR signal triggered at $t_0$, and the hands are detected to be put on the steering wheel at $t_1$. Under the haptic guidance, the driver intervenes the control at $t_2$. Once driver's state is considered as fully qualified for manual driving at $t_4$, the haptic assistance as well as the contribution of automation are removed, and the takeover process is completed.

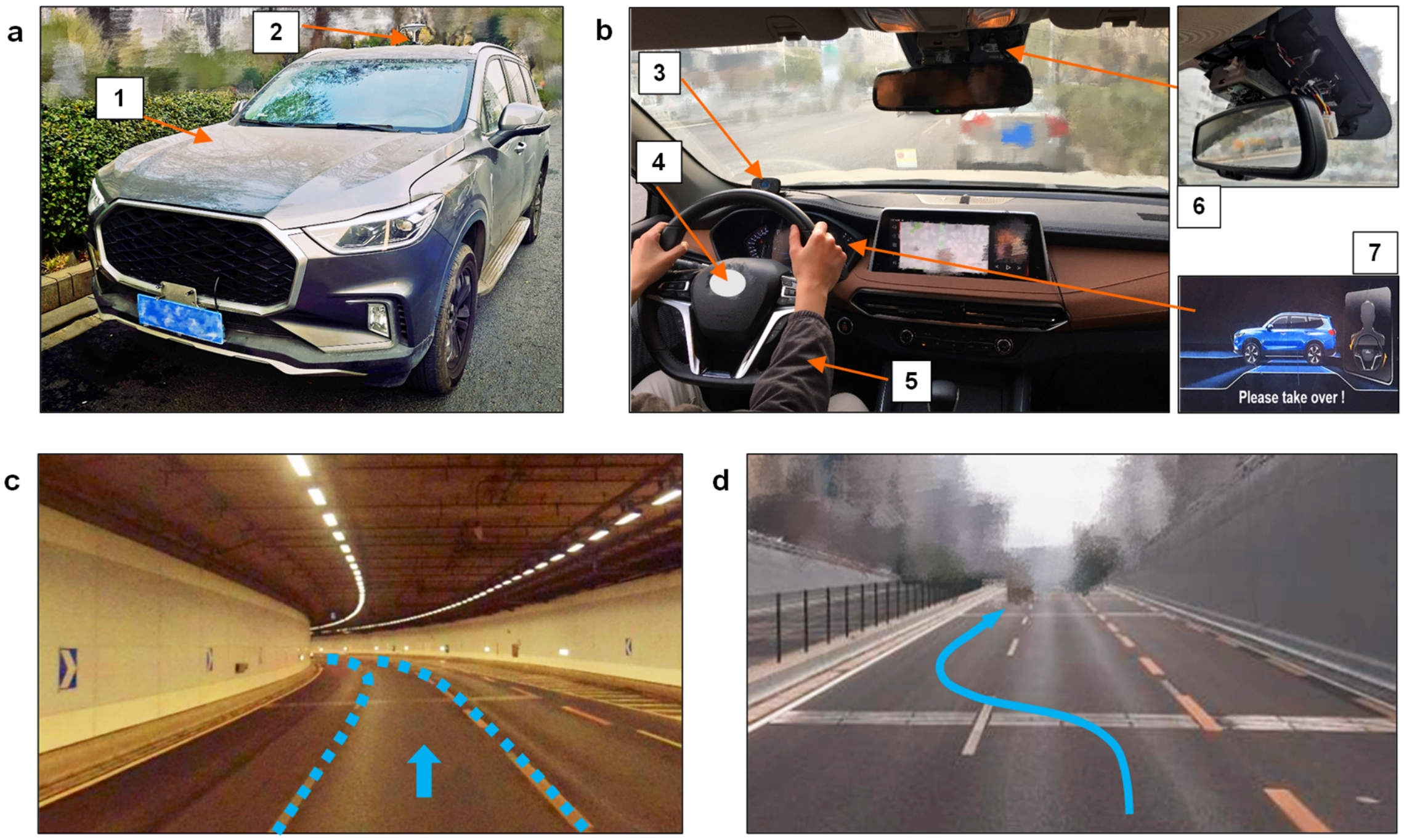

1: The experimental vehicle; 2: GPS antenna; 3: Driver state monitoring system; 4: The haptic steering system; 5: The participant; 6: The perception system of the automated vehicle; 7: TOR signal (visual part).

**Fig. 3. Experimental set-up. a**, The experimental automated vehicle used in this study was a sport utility vehicle. **b**, View from inside the testing vehicle. Key components used in the experiment include the driver state monitoring system, the haptic steering system, the perception system demanded for lane detection of automated vehicle, and the TOR signal. **c**, The scenario of task A, which was set as the automation-to-driver takeover control during a normal steering maneuver under a single-lane condition. **d**, The scenario of task B, which was designed as automation-to-driver takeover control under a single lane change maneuver.

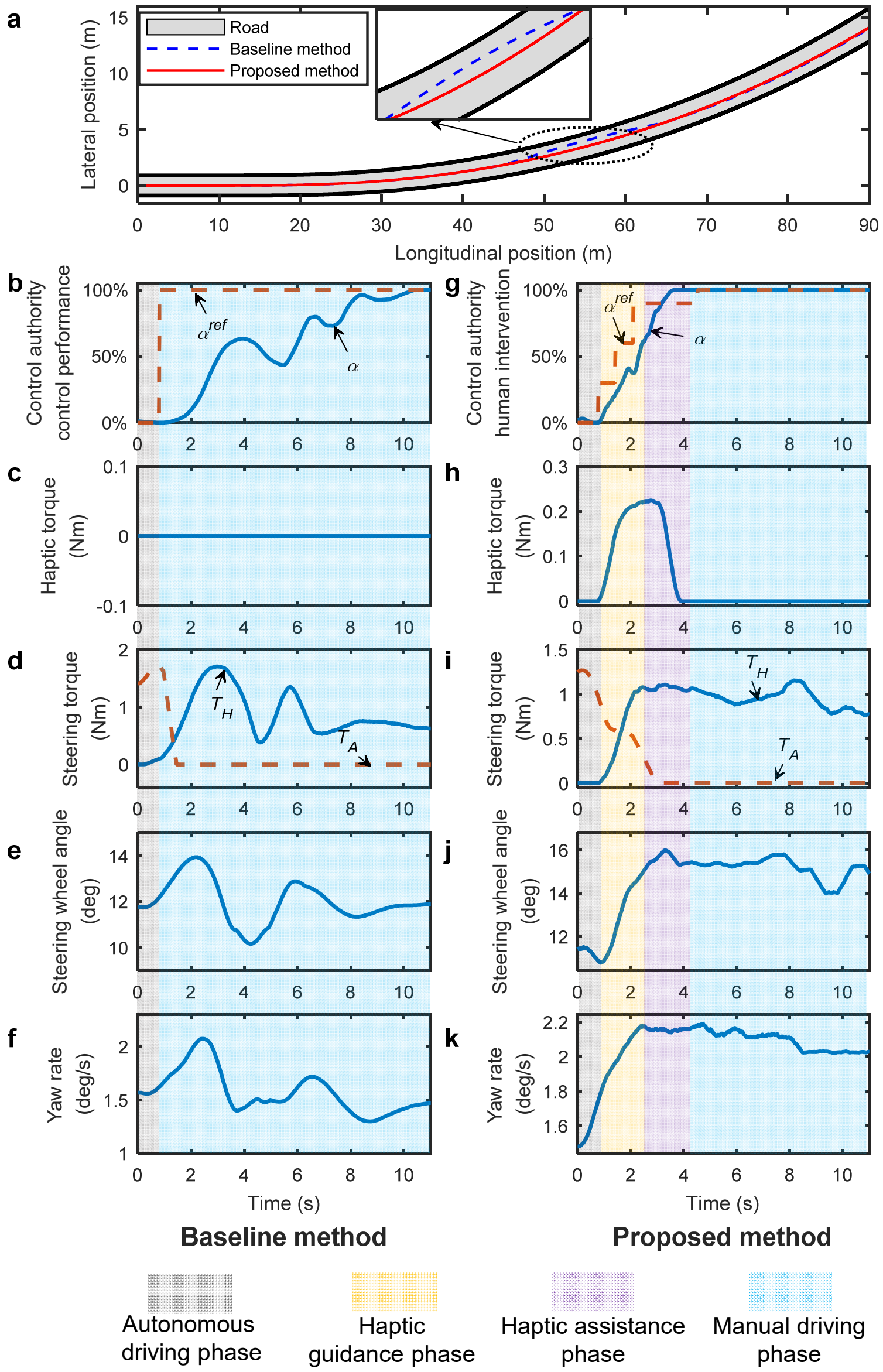

**Fig. 4. Example data for a representative participant while conducting Task A with baseline and proposed methods. a**, The paths of the vehicle during the takeover trails with the participant under the baseline and the proposed methods. **b**, Allowed driver authority, and the control performance for driving with baseline method. **c** and **h**, Haptic torque versus time during each trail. There was no haptic torque applied during takeover with the baseline method. Under the proposed method, the haptic steering torque was generated. **d** and **i**, Driver's steering torque and the contribution from automation versus time during each trail. Under the baseline method, the participant made many oscillations in the steering torque to handle the vehicle. With the proposed method, the driver smoothly applied the steering torque under the haptic interface. **e** and **j**, Steering wheel angle versus time during each trail. **g**, Allowed driver authority, and the degree of driver's intervention versus time during the trail with proposed method. **f** and **k**, The yaw rate of the vehicle versus time during each trail.

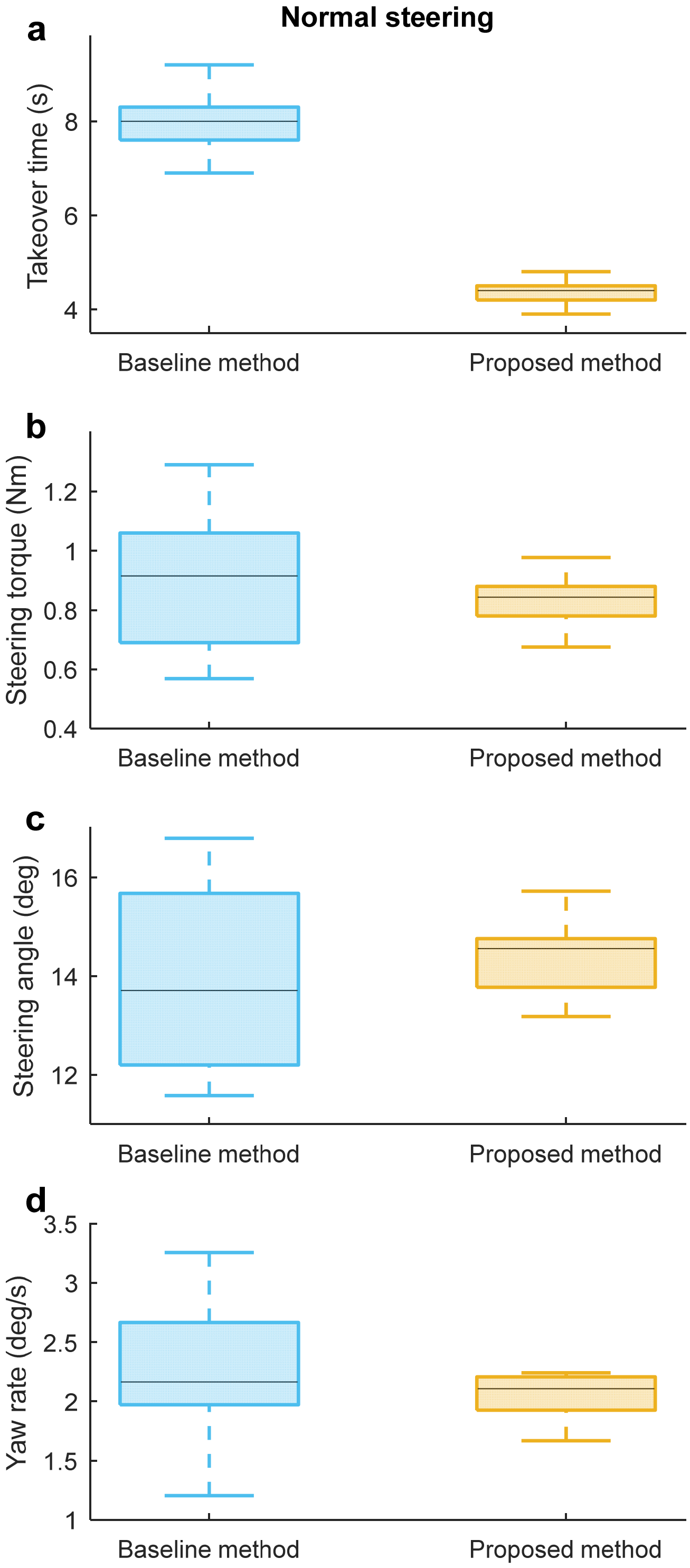

**Fig. 5. Box plots of the experimental results of Task A. a**, Results of takeover time. **b**, Results of steering torque. **c**, Results of steering wheel angle. **d**, Results of yaw rate of the vehicle.

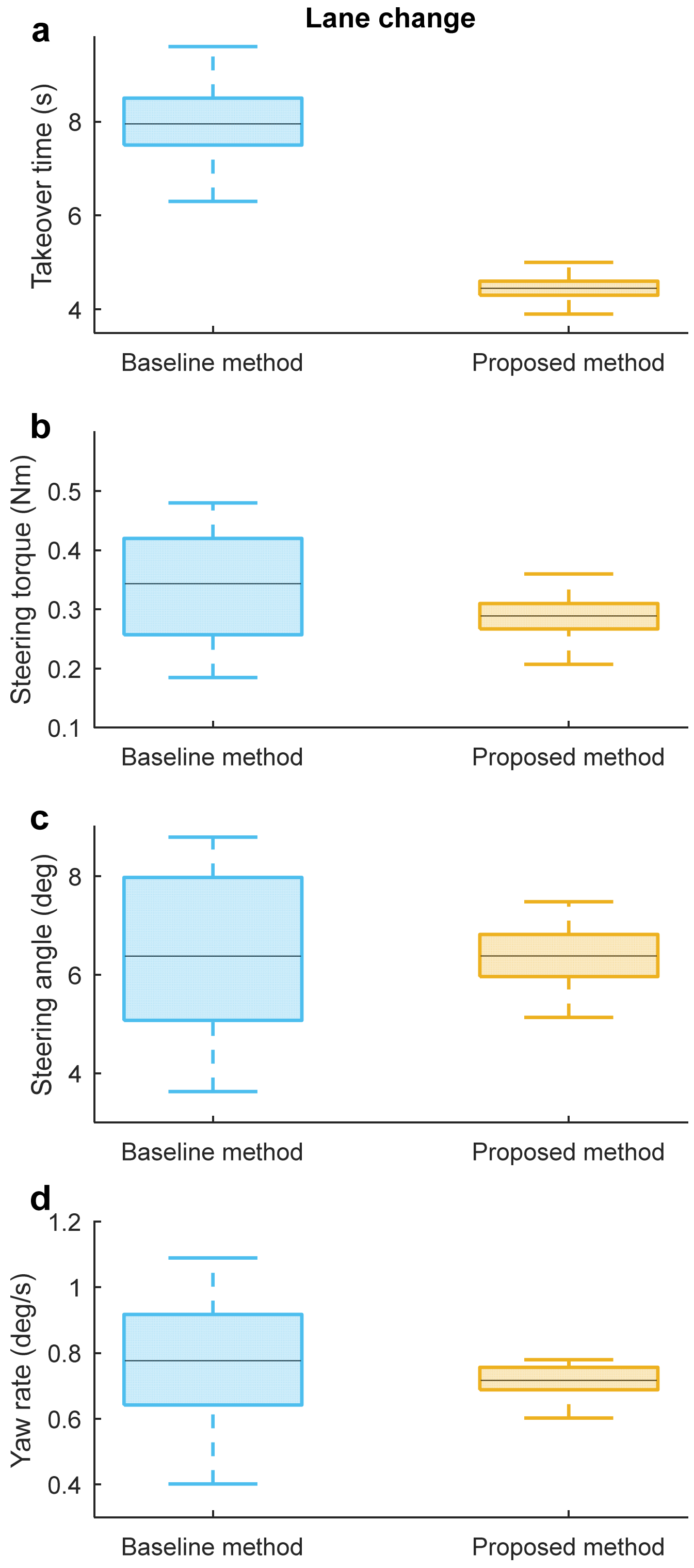

**Fig. 6. Box plots of the experimental results of Task B. a**, Results of takeover time. **b**, Results of driver steering torque. **c**, Results of steering wheel angle. **d**, Results of yaw rate of the vehicle.

Supplementary Materials for

# Human-Machine Collaboration for Automated Vehicles via an Intelligent Two-Phase Haptic Interface

**Authors**

Chen Lv[1]*, Yutong Li[1], Yang Xing[1], Chao Huang[1], Dongpu Cao[2], Yifan Zhao[3], Yahui Liu[4]

**Affiliations**

1. School of Mechanical and Aerospace Engineering, Nanyang Technological University, 50 Nanyang Ave, 639798, Singapore
2. Mechanical and Mechatronics Engineering, University of Waterloo, ON, N2L 3G1, Canada
3. School of Aerospace, Transport and Manufacturing, Cranfield University, Bedford, MK43 0AL, UK
4. School of Vehicle and Mobility, Tsinghua University, Beijing 100084, China

*Corresponding author. Email: lyuchen@ntu.edu.sg.

**Supplementary Note 1**

Experimental vehicle

The experimental sport utility vehicle shown in Fig. 3a was modified and used as the testing platform for a range of experiments in automated driving. The powertrain system of the vehicle consisted of a 100 kW internal combustion engine with a peak torque of 200 N·m, and an automatic gearbox. In order to realize the functionality of autonomous driving, the actuator employed for steering was modified from a conventional electronic power steering (EPS) system. The modified steering system consists of two DC electric motors, a gear reduction, and an electric control unit that communicates with the vehicle controller via a controller area network (CAN) bus. This dual-motor set-up was to physically separate the haptic torque and automation's torque. The command of the haptic torque $T_{hpt}$ was received by a DC motor mounted under the steering wheel in the EPS system. Thus, the haptic torque $T_{hpt}$ was applied directly on the steering wheel, and it would be felt by the human driver, resulting in driver's actual torque input $T_H$. While the command signal of the automation's torque $T_A$ was received by the other DC motor, which was additionally added at the far end of the steering column. Therefore, the automation's torque was applied directly to the downstream plant rather than the steering wheel. The maximum steering angle of the front wheels was ±20°. The torque input by the driver and the angle of the steering wheel were measured by torque and angular position sensors mounted in the EPS system. The accuracy of the steering torque measurement is within 3%. Some of the key parameters of the experimental vehicle are listed in Supplementary Table 3.

The localization and dynamic state measurements of the vehicle were respectively provided by a global navigation satellite system (GNSS) receiver and an inertial measurement unit. The available measurements including vehicle position, longitudinal and lateral velocities, longitudinal and lateral accelerations, and yaw angle and yaw rate, were obtained with a sampling frequency of 100 Hz. The accuracy of the yaw rate measurement is within 5%. A differential global positioning system was employed to augment the GNSS measurements for ensuring a positioning error of less than 10 cm. A driver monitoring system (HiRain Technologies) was able to detect the in-vehicle activities of the human driver from video imagery obtained with an onboard camera, and thereby estimate the driver's level of attention to the driving task.

All perception, sensing, planning, and control algorithms were run on an onboard real-time computational platform MicroAutoBox II (dSPACE GmbH). The control algorithm was developed in Matlab/Simulink (version R2017a, MathWorks), and updated at a frequency of 50 Hz. The MPC algorithm was formulated as a quadratic programming problem and solved by the toolbox of qpOASES (*31*). The execution of C code was automatically generated by Matlab/Simulink, and data acquisition was conducted using ControlDesk software (dSPACE GmbH). The recorded data were exported to Matlab and filtered by using the moving-average filter (version 2017b, MathWorks) for further analysis.

**Supplementary Note 2**

Method for computing the optimal system input

The optimal system input $T_{ref}$ which ensures that the vehicle tracks the lane centerline can be calculated by solving the following formulated optimization problem with MPC (*38, 39*):

$$\min_{\mathbf{u}_k} \sum_{i=1}^{N} \left\| r_{i|k} - r_{i|k}^{ref} \right\|_{W_1}^2 + \sum_{i=0}^{N-1} \left\| T_{ref,i|k} \right\|_{Q_1}^2 \tag{1a}$$

$$\text{s.t. } x_{i+1|k} = g_d(x_{i|k}, u_{i|k}) \tag{1b}$$

$$r_{i|k} = Cx_{i|k} \tag{1c}$$

$$T_{ref,\min} \le T_{ref,i|k} \le T_{ref,\max} \tag{1d}$$

$$\Delta T_{ref,\min} \le \Delta T_{ref,i|k} \le \Delta T_{ref,\max} \tag{1e}$$

$$x_{0|k} = x_k,\ T_{ref,-1|k} = T_{ref,k-1} \tag{1f}$$

where the output matrix $C$ = [1, 0, 0, 0, 0, 0, 0; 0, 0, 0, 1, 0, 0, 0], $N$ is the prediction horizon, $W_1$ and $Q_1$ are weighting factors. The dynamics model $g_d$ in supplementary equation (1b) is obtained by discretizing supplementary equation (2) using the Euler method. $x_{i|k}$ denotes the $i$th state prediction at time step $k$ obtained by applying the optimal input sequence $\mathbf{u}_k = \{T_{ref,0|k}, T_{ref,1|k}, \dots, T_{ref,N-1|k}\}$ to the discrete-time vehicle dynamics model shown in supplementary equation (1b), starting from the state in supplementary equation (1f) measured at current time step. The constraints are described in supplementary equations (1d) and (1e), respectively. The value selections of the key parameters used in the MPC are listed in Supplementary Table 5.

The dynamic model adopted for describing the continuous behaviors of the vehicle can be described as (*40*):

$$\dot{x} = g(x, u) \tag{2}$$

where

$$g = \begin{pmatrix} v_x \sin\psi + \dot{y}\cos\psi \\ \dot{y} \\ a_1\dot{y} + a_2\dot{\psi} + a_3\delta \\ \dot{\psi} \\ a_4\dot{y} + a_5\dot{\psi} + a_6\delta \\ \dot{\delta} \\ a_7\dot{y} + a_8\dot{\psi} + a_9\delta + a_{10}\dot{\delta} + u/I_{sw} \end{pmatrix},\ x = \begin{bmatrix} Y \\ y \\ \dot{y} \\ \psi \\ \dot{\psi} \\ \theta_{sw} \\ \dot{\theta}_{sw} \end{bmatrix},\ u = T_{ref},\ a_1 = \frac{K_f + K_r}{mv_x},\ a_2 = \frac{K_f l_f - K_r l_r}{mv_x} + v_x,\ a_3 = -\frac{K_f}{mi_{sw}},$$

$$a_4 = \frac{K_f l_f - K_r l_r}{I_z v_x},\ a_5 = \frac{K_f l_f^2 + K_r l_r^2}{I_z v_x},\ a_6 = -\frac{K_f l_f}{I_z i_{sw}},\ a_7 = \frac{K_{alig}}{I_{sw} v_x i_{sw}},\ a_8 = \frac{K_{alig} l_f}{I_{sw} v_x i_{sw}},\ a_9 = -\frac{K_{sw}}{I_{sw}},\ a_{10} = -\frac{B_{sw}}{I_{sw}}.$$

The $Y$ and $y$ denote the lateral position of the center of vehicle gravity in global and body frame (Supplementary Fig. 7). The yaw angle $\Psi$ denotes the rotation of center of gravity around $z$-axis. The vehicle mass, moment of inertia and distances from the center of gravity to front and rear axles are denoted as $m$, $I_z$, $l_f$ and $l_r$, respectively. The cornering stiffness of the front and rear axles are denoted as $K_f$ and $K_r$, respectively. $I_{sw}$, $B_{sw}$ and $K_{sw}$ are the equivalent moment of inertia, damping and stiffness coefficients of the steering system, respectively. $i_{sw}$ is the transmission ratio of the steering system. The value selection of the key parameters of the vehicle model is listed in Supplementary Table 3.

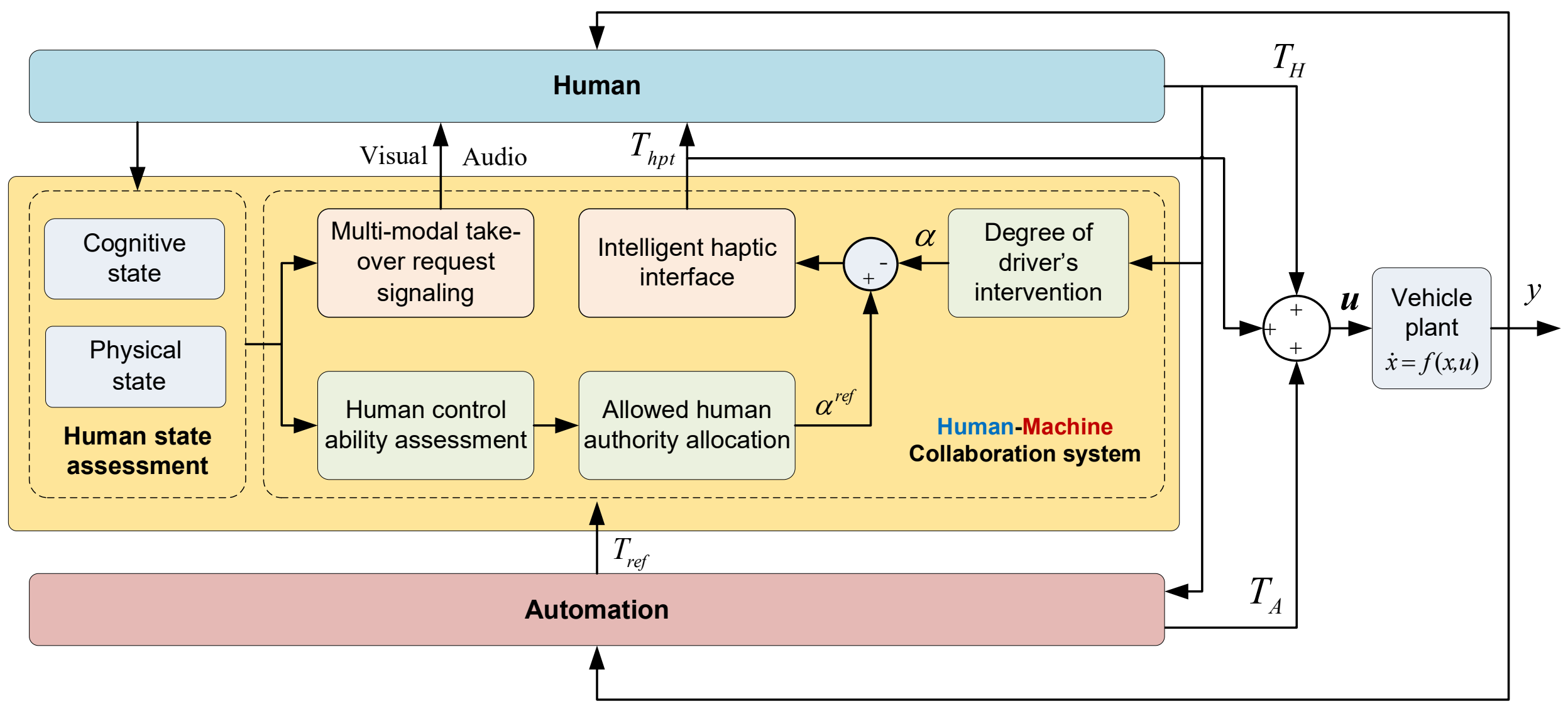

**Supplementary Fig. 1. Control block diagram of the proposed human-automation collaboration system with the intelligent haptic interface.** During the takeover process, an optimal sequence of control input $T_{ref}$ will firstly be derived from the planned trajectory of the automated vehicle. In the meantime, the developed human-automation collaboration system assesses the driver's states and control ability in real time, and then decides how much control authority could be allocated to human driver, gradually increasing this with the recovery of driver's states and control performance. To do this, a human authority allocation module was designed to calculate the allowed driver authority $\alpha^{ref}$ in real time, based upon the driver's cognitive attention, neuromuscular state and the required driving task. The real value of the driver's degree of intervention $\alpha$ will then be compared to the allowed one $\alpha^{ref}$. The intelligent haptic feedback torque will then be generated on the steering wheel, in order to minimize the deviation between $\alpha$ and $\alpha^{ref}$. The haptic steering torque applied is expected to guide or assist the driver to use appropriate degree of steering torque in an appropriate manner so as to gradually complete the overall handover process. Thus, during the automation-human takeover transition, the human and the machine dynamically share the control authority, jointly completing the required driving task. The steering torque contributed by the human driver takes up $\alpha$ percent of the overall torque applied to the vehicle. And the input contributed by automation system $T_A$, always compensates for the summation of driver's actual torque and the haptic torque, occupying the remaining part of the optimal control input. Once the $\alpha$ increases to *100%*, then it indicates the takeover process has been completed. The detailed information of each module within this framework is described in each corresponding section of Methods in the main text.

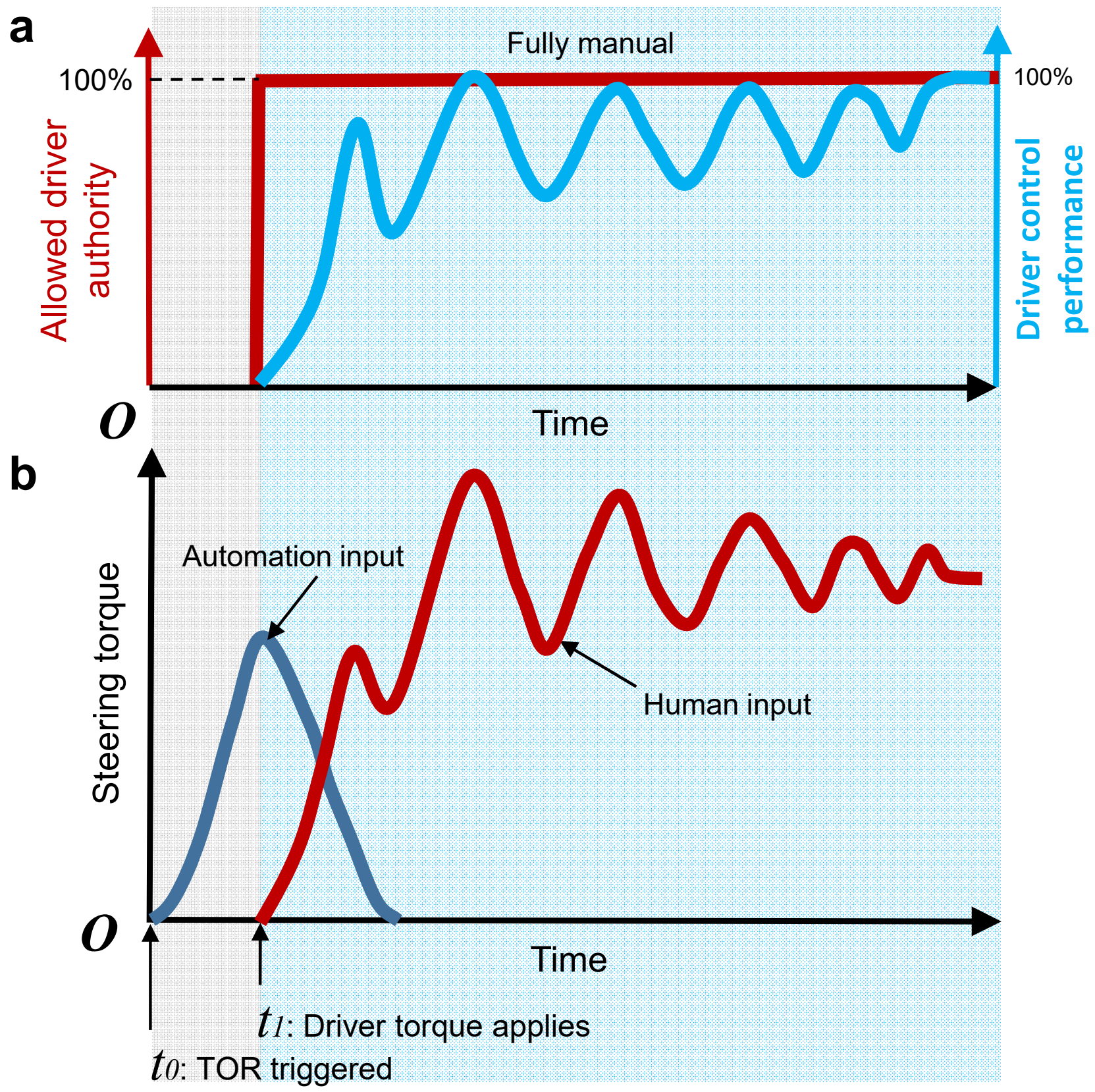

**Supplementary Fig. 2. Schematic diagram of the takeover process with baseline strategy, which a fade out of autopilot steering torque. a**, The allowed control authority and assessed driver's driving performance. The autopilot steering torque is decreased with a fixed slope at slope of 2.5 N·m/s once the driver intervenes to the control at $t_1$. **b**, Perceiving the takeover request signal triggered at $t_0$, the human driver disengages from the pre-occupied NDA and transitions to the driving task. At $t_1$, the driver intervenes the control with steering torque applied to the hand wheel.

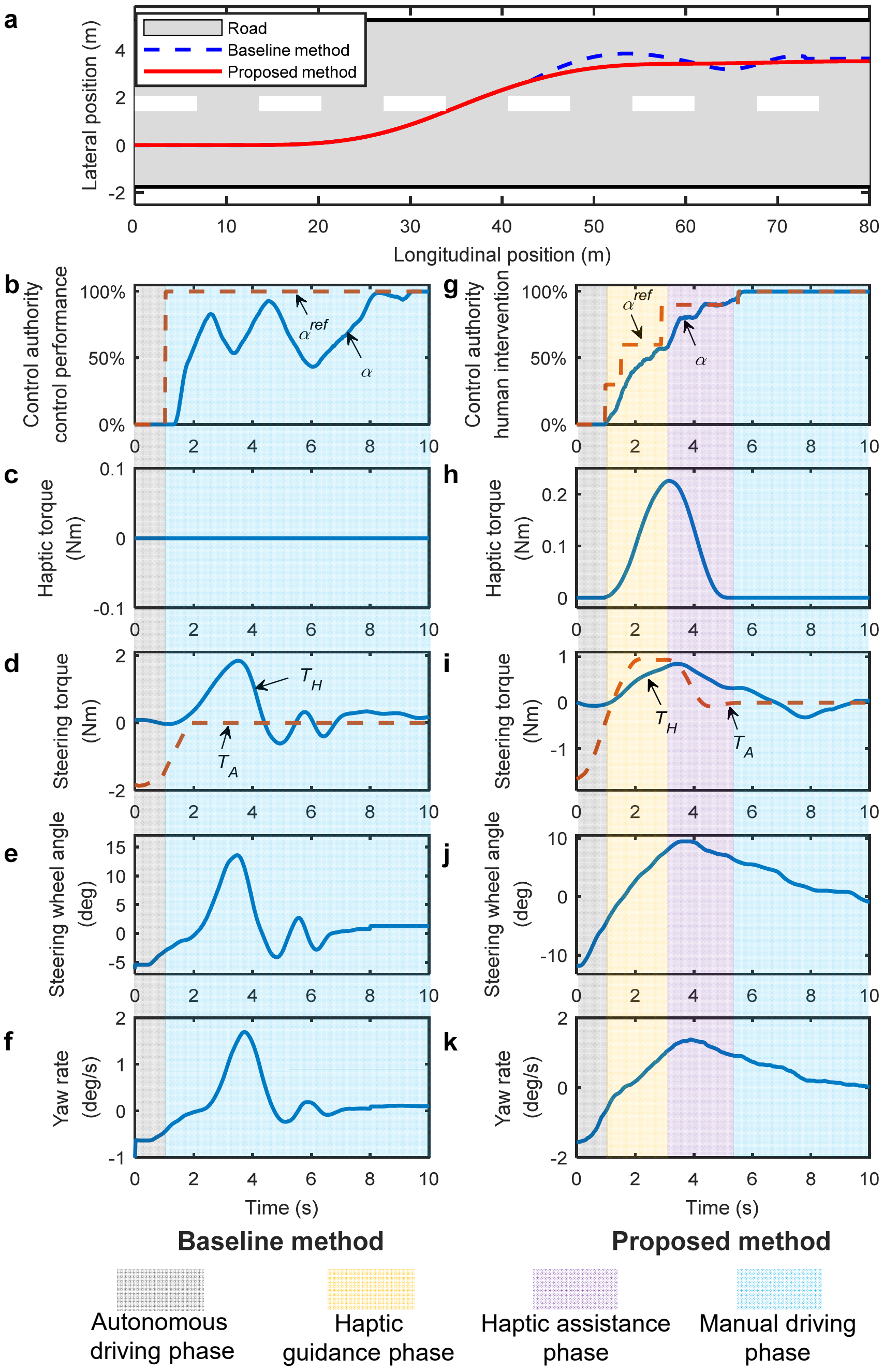

**Supplementary Fig. 3. Example data for a representative participant while conducting Task B. a**, The paths of the vehicle during the takeover trails with the participant under the proposed method (red) and under the baseline method (dashed blue). **b**, Allowed driver authority $\alpha^{ref}$ (dashed red), and driver's control performance $\alpha$ in baseline group. **c** and **h**, Haptic steering torque versus time during each trail of Task B. **d** and **i**, Driver's steering torque (solid blue) and the contribution from automation system (dashed red) versus time during each trail of Task B. **g**, Allowed driver authority $\alpha^{ref}$ (dashed red), and the actual degree of driver's intervention $\alpha$ in test with proposed method versus time. **e** and **j**, The steering wheel angle versus time during each trail of Task B. **f** and **k**, The yaw rate of the vehicle versus time during each trail of Task B.

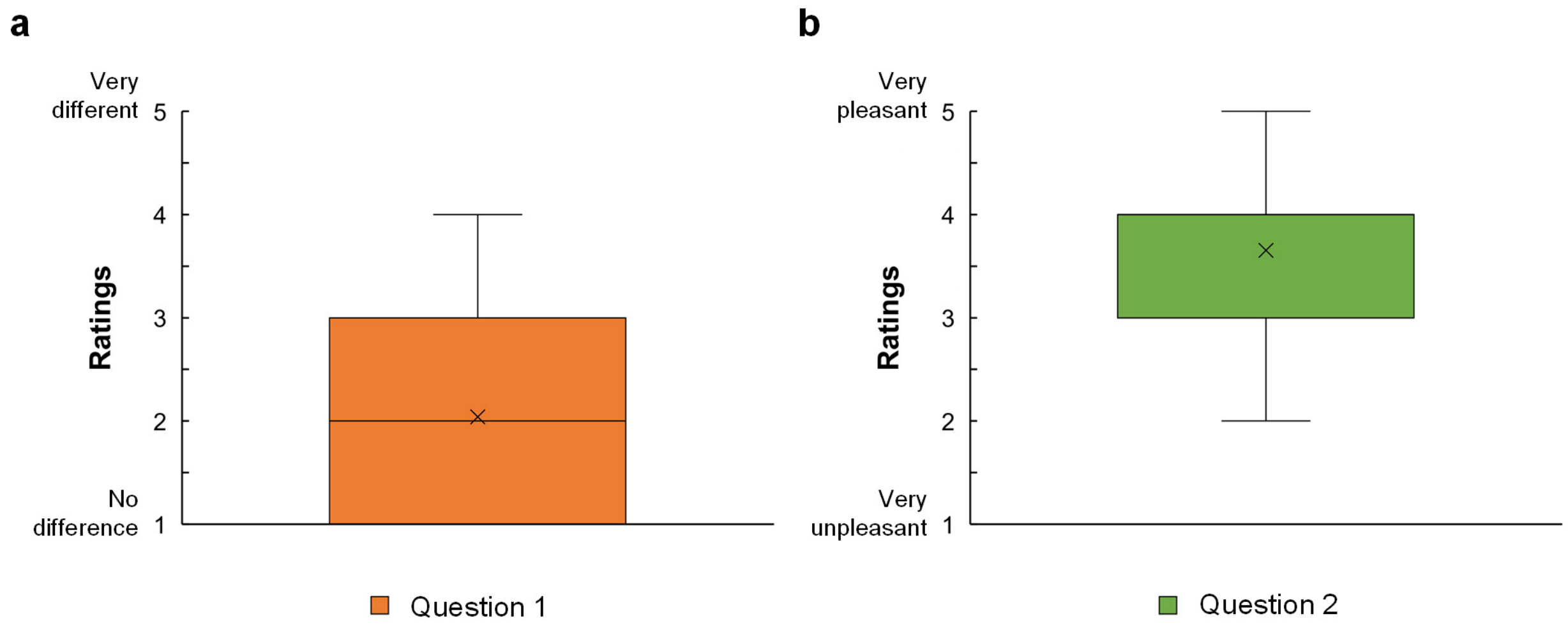

**Supplementary Fig. 4. Evaluation of the responses to the questions. a**, Results of the responses to the Question 1. In terms of the steering feeling, a slight difference between the two methods was rated with mean value of 2.0 ± 1.06 (rating scale from 1: no difference to 5: very different). **b**, Results of the responses to the Question 2. The pleasantness of the steering feeling under the proposed approach was rated at 3.7 ± 0.62 (rating scale from 1: very unpleasant to 5: very pleasant).

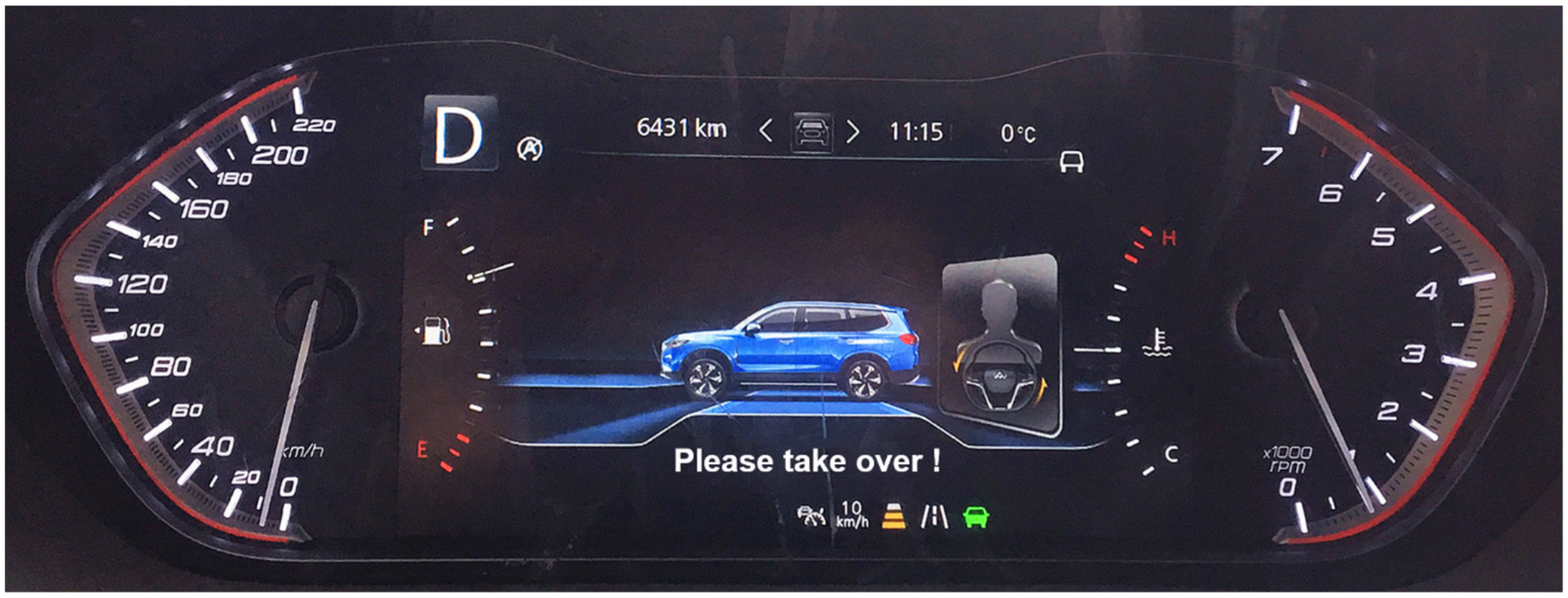

**Supplementary Fig. 5. The visual takeover request signal.** When the multi-modal takeover request signal was triggered, visual request signal of the text "Please take over!" was shown on the dashboard.

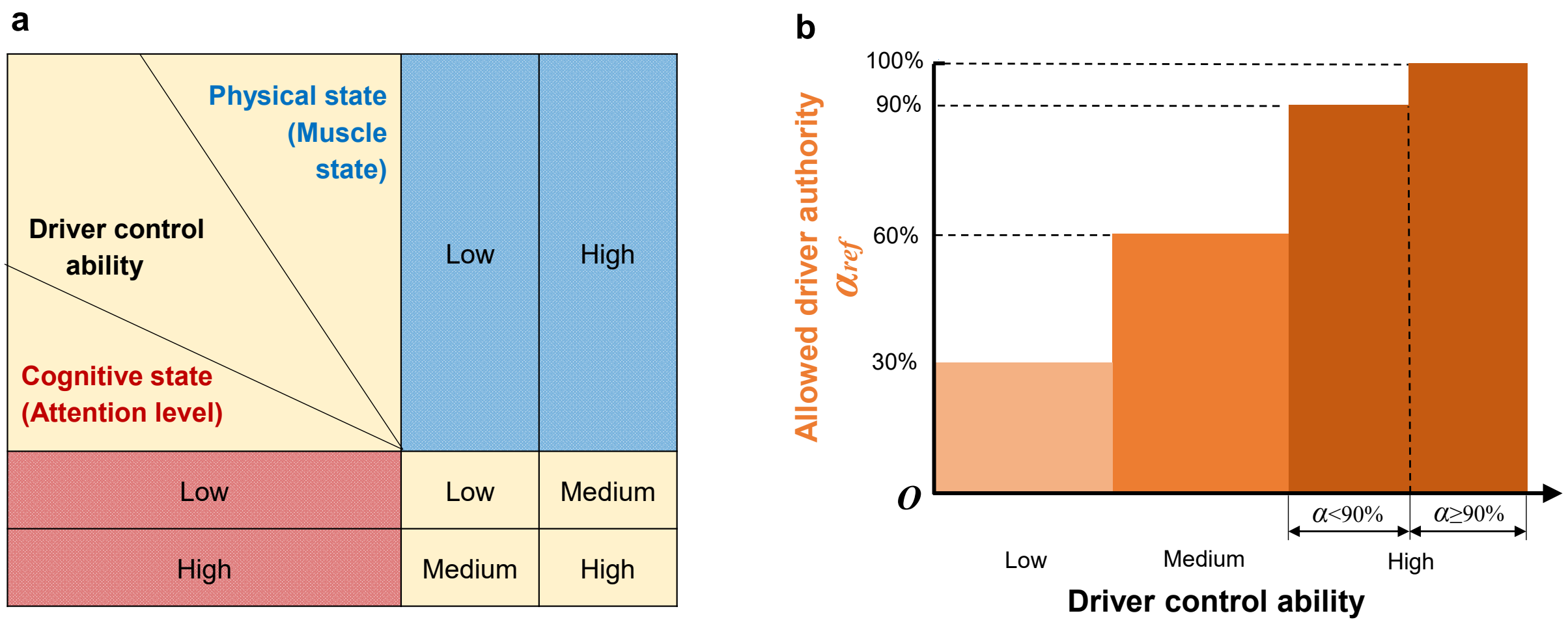

**Supplementary Fig. 6. The assessment and decision mechanisms for driver control ability and authority. a**, Driver's control ability is assessed based upon the attention level and muscle state. For cognitive state assessment, the onboard driver monitoring system detects the eye gaze direction of the driver as well as the driver's current behavior (i.e., driving or non-driving), and thereby assesses the current level (high or low) of the driver's attention to the driving activity. For physical state assessment, muscle stiffness coefficient K is selected as the key indicator. The level of muscle state is considered high when $K$ exceeds a predefined threshold $K_1$. Otherwise, the driver's physical state is considered to be low. When both the cognitive and physical states of the driver are low, the driver control ability is considered as low. When only one of the two states is low, the control ability is considered as medium. And when both the cognitive and physical states of the driver are high, then the driver control ability is considered as high. **b**, The maximum allowed driver control authority $\alpha^{ref}$ is decided based on driver's control ability. When driver's control ability is considered as low, then the value of $\alpha^{ref}$ is set as 30%. When driver's control ability is considered as medium, then the value of $\alpha^{ref}$ is set as 60%. When driver's control ability is considered as high, and the degree of driver's intervention $\alpha$ is below 90%, then the value of $\alpha^{ref}$ is set as 90%. When driver's control ability is considered as high, and the degree of driver's intervention $\alpha$ is 90% and above, then the value of $\alpha^{ref}$ is set as 100%. The detailed decision mechanism of $\alpha^{ref}$ can be also expressed by supplementary equation (3).

$$\alpha^{ref} = \begin{cases} 30\% & (\text{If} \quad \text{driver control ability==Low}) \\ 60\% & (\text{If} \quad \text{driver control ability==Medium}) \\ 90\% & (\text{If} \quad \text{driver control ability==High} \quad \& \quad \alpha < 90\%) \\ 100\% & (\text{If} \quad \text{driver control ability==High} \quad \& \quad 90\% \le \alpha \le 100\%) \end{cases} \tag{3}$$

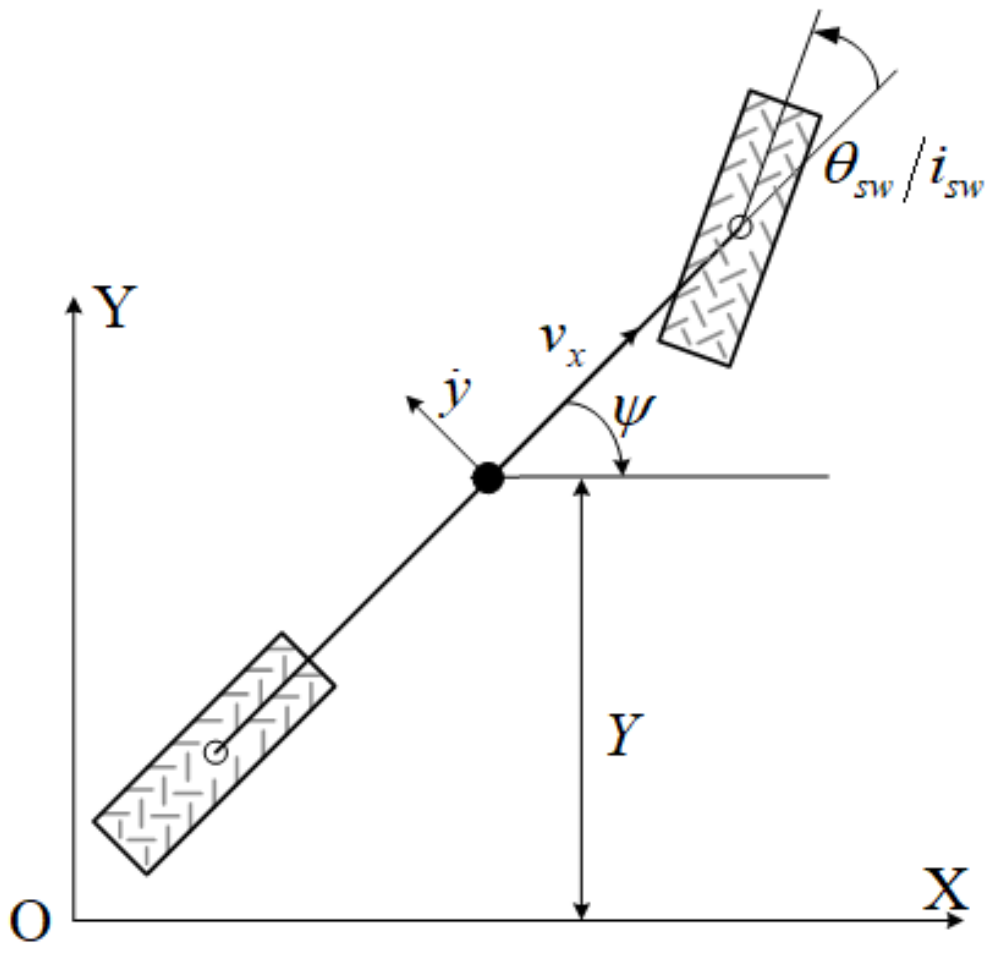

**Supplementary Fig. 7. The bicycle model used for describing vehicle dynamic behaviors.** The lateral position of the center of gravity (CoG) in global and body frame are denoted as $Y$ and $y$, respectively. The yaw angle $\Psi$ denotes the rotation of CoG around z-axis, $\theta_{sw}$ is the steering angle of the hand wheel, and $i_{sw}$ is the ratio of gear reduction in the steering system. In this paper we only focus on the takeover control during steering maneuver, thus the vehicle is assumed to travel at a constant longitudinal velocity $v_x$. The definition and value selection of the key parameters of the vehicle model are listed in Supplementary Table 3.

**Supplementary Table 1.**

Statistical analysis of the experimental data for each participant in Task A.

| Task A: Normal steering | | | | | | | | | | | | | | | |
|---|---|---|---|---|---|---|---|---|---|---|---|---|---|---|---|
| Participant | Takeover time (s) | | Steering wheel torque (Nm) | | | | Steering wheel angle (deg) | | | | Yaw rate (deg/s) | | | | |
| | Mean | | Mean | | S. D. | | Mean | | S. D. | | Mean | | S. D. | | |
| | Baseline | Proposed | Baseline | Proposed | Baseline | Proposed | Baseline | Proposed | Baseline | Proposed | Baseline | Proposed | Baseline | Proposed | |
| 1 | 8 | 4.2 | 1.13 | 0.78 | 0.86 | 0.42 | 12.12 | 14.67 | 5.09 | 1.73 | 3.25 | 1.66 | 0.93 | 0.32 | |
| 2 | 7.3 | 4.5 | 0.75 | 0.89 | 0.36 | 0.41 | 15.86 | 13.19 | 5.68 | 1.57 | 1.70 | 2.20 | 1.06 | 0.29 | |
| 3 | 8.1 | 4.1 | 0.66 | 0.86 | 0.65 | 0.35 | 15.94 | 15.33 | 4.94 | 1.96 | 2.23 | 2.23 | 0.86 | 0.29 | |
| 4 | 7.6 | 4.4 | 1.06 | 0.83 | 0.87 | 0.42 | 11.58 | 14.64 | 4.99 | 1.78 | 3.15 | 1.84 | 1.32 | 0.28 | |
| 5 | 7.8 | 4.5 | 1.29 | 0.89 | 0.86 | 0.32 | 12.94 | 14.93 | 2.49 | 1.77 | 2.49 | 2.21 | 1.05 | 0.28 | |
| 6 | 8.8 | 4.5 | 0.88 | 0.84 | 0.44 | 0.31 | 12.18 | 14.86 | 4.33 | 1.78 | 1.93 | 2.23 | 1.19 | 0.30 | |
| 7 | 7.5 | 4.3 | 0.56 | 0.78 | 0.37 | 0.37 | 13.63 | 14.71 | 4.43 | 1.78 | 2.04 | 2.24 | 1.59 | 0.30 | |
| 8 | 7.8 | 4.6 | 0.69 | 0.88 | 0.32 | 0.33 | 12.78 | 13.64 | 1.44 | 1.72 | 2.74 | 2.16 | 0.90 | 0.36 | |
| 9 | 8.3 | 4.4 | 0.64 | 0.85 | 0.34 | 0.40 | 11.71 | 13.77 | 5.42 | 1.75 | 2.29 | 2.20 | 1.00 | 0.27 | |
| 10 | 6.9 | 4.5 | 0.86 | 0.67 | 0.67 | 0.35 | 13.38 | 13.74 | 5.65 | 1.65 | 2.63 | 1.73 | 1.23 | 0.32 | |
| 11 | 7.7 | 4.6 | 1.16 | 0.75 | 0.76 | 0.32 | 14.58 | 13.79 | 5.11 | 1.64 | 2.76 | 2.17 | 0.40 | 0.28 | |
| 12 | 9.2 | 4.8 | 0.96 | 0.97 | 0.94 | 0.38 | 14.67 | 14.57 | 2.67 | 1.79 | 1.84 | 1.83 | 0.98 | 0.29 | |
| 13 | 8.3 | 4.5 | 0.66 | 0.86 | 0.35 | 0.32 | 13.12 | 13.71 | 3.99 | 1.57 | 2.16 | 1.94 | 1.17 | 0.29 | |
| 14 | 8 | 4.3 | 0.89 | 0.76 | 0.62 | 0.38 | 11.95 | 14.28 | 5.59 | 1.77 | 2.69 | 2.24 | 1.88 | 0.32 | |
| 15 | 8.1 | 4.7 | 0.75 | 0.83 | 0.35 | 0.31 | 16.65 | 13.73 | 4.86 | 1.61 | 1.66 | 1.83 | 1.86 | 0.28 | |
| 16 | 8.1 | 4.4 | 0.94 | 0.96 | 0.77 | 0.44 | 13.45 | 14.16 | 5.34 | 1.68 | 2.87 | 1.97 | 1.09 | 0.36 | |
| 17 | 9.4 | 4.5 | 0.96 | 0.82 | 0.86 | 0.29 | 14.55 | 15.05 | 4.98 | 1.83 | 1.20 | 1.96 | 0.95 | 0.32 | |
| 18 | 8.8 | 4.1 | 1.29 | 0.79 | 0.85 | 0.44 | 11.98 | 14.66 | 5.66 | 1.85 | 1.97 | 2.21 | 1.02 | 0.28 | |
| 19 | 7.6 | 4.6 | 1.03 | 0.91 | 0.78 | 0.33 | 13.79 | 13.24 | 4.99 | 1.77 | 2.15 | 2.08 | 1.14 | 0.32 | |
| 20 | 8 | 4.2 | 0.69 | 0.75 | 0.30 | 0.45 | 16.59 | 15.42 | 1.99 | 1.88 | 2.16 | 2.12 | 0.21 | 0.31 | |
| 21 | 8.2 | 4.5 | 1.19 | 0.87 | 0.66 | 0.35 | 15.63 | 14.61 | 4.55 | 1.77 | 1.98 | 2.17 | 1.10 | 0.30 | |
| 22 | 7.6 | 4.2 | 1 | 0.84 | 0.56 | 0.39 | 11.95 | 13.76 | 2.22 | 1.78 | 1.57 | 2.04 | 0.99 | 0.34 | |
| 23 | 7.1 | 4.2 | 1.09 | 0.67 | 0.73 | 0.34 | 16.79 | 15.72 | 5.45 | 1.97 | 2.66 | 2.06 | 1.00 | 0.20 | |
| 24 | 7.9 | 4.2 | 0.75 | 0.88 | 0.46 | 0.39 | 15.61 | 14.75 | 5.43 | 1.35 | 2.09 | 2.14 | 0.95 | 0.26 | |
| 25 | 8 | 4.3 | 0.96 | 0.70 | 0.62 | 0.35 | 13.89 | 14.58 | 5.33 | 1.66 | 2.14 | 1.92 | 1.06 | 0.29 | |
| 26 | 8.7 | 3.9 | 0.68 | 0.87 | 0.33 | 0.47 | 15.72 | 13.75 | 5.83 | 1.64 | 2.52 | 1.81 | 1.03 | 0.29 | |

**Supplementary Table 2.**

Statistical analysis of the experimental data for each participant in Task B.

| Task B: Lane change | | | | | | | | | | | | | | |
|---|---|---|---|---|---|---|---|---|---|---|---|---|---|---|
| Participant | Takeover time (s) | | Steering wheel torque (Nm) | | | | Steering wheel angle (deg) | | | | Yaw rate (deg/s) | | | |
| | Mean | | Mean | | S. D. | | Mean | | S. D. | | Mean | | S. D. | |
| | Baseline | Proposed | Baseline | Proposed | Baseline | Proposed | Baseline | Proposed | Baseline | Proposed | Baseline | Proposed | Baseline | Proposed |
| 1 | 7.9 | 4.8 | 0.22 | 0.28 | 0.76 | 0.28 | 8.73 | 6.96 | 5.68 | 3.33 | 0.85 | 0.77 | 1.71 | 1.13 |
| 2 | 8 | 5 | 0.27 | 0.26 | 0.73 | 0.33 | 4.96 | 6.22 | 2.41 | 2.62 | 1.08 | 0.66 | 1.30 | 0.85 |
| 3 | 7.7 | 4.5 | 0.40 | 0.29 | 0.41 | 0.38 | 8.57 | 5.96 | 5.47 | 3.39 | 0.68 | 0.70 | 0.85 | 0.91 |
| 4 | 8.2 | 4.4 | 0.36 | 0.33 | 0.80 | 0.32 | 7.36 | 6.79 | 4.99 | 3.12 | 0.74 | 0.76 | 0.67 | 1.26 |
| 5 | 6.8 | 4.5 | 0.43 | 0.31 | 1.03 | 0.43 | 6.31 | 7.16 | 2.18 | 3.32 | 1.37 | 0.64 | 1.02 | 0.99 |
| 6 | 9 | 4.1 | 0.22 | 0.26 | 0.77 | 0.38 | 6.74 | 6.44 | 6.5 | 3.43 | 1.60 | 0.65 | 1.04 | 0.80 |
| 7 | 7.1 | 4.7 | 0.35 | 0.24 | 0.45 | 0.38 | 5.58 | 7.16 | 5.86 | 3.32 | 0.17 | 0.77 | 1.05 | 0.57 |
| 8 | 7.5 | 4.6 | 0.47 | 0.29 | 0.49 | 0.39 | 7.64 | 5.14 | 5.69 | 3.24 | 0.64 | 0.71 | 0.58 | 0.74 |
| 9 | 8.6 | 4.3 | 0.26 | 0.29 | 0.96 | 0.25 | 7.93 | 6.35 | 5.94 | 3.36 | 0.59 | 0.85 | 1.16 | 0.73 |
| 10 | 7.6 | 4.6 | 0.31 | 0.33 | 0.94 | 0.27 | 6.49 | 5.83 | 1.19 | 3.49 | 0.55 | 0.70 | 2.13 | 0.93 |
| 11 | 8.5 | 4.4 | 0.42 | 0.28 | 0.42 | 0.36 | 8.03 | 5.58 | 5.55 | 3.71 | 0.91 | 0.72 | 1.15 | 1.20 |
| 12 | 8 | 4.1 | 0.36 | 0.26 | 0.77 | 0.42 | 5.66 | 6.72 | 5.87 | 3.12 | 1.03 | 0.71 | 1.36 | 0.86 |
| 13 | 9.6 | 4.3 | 0.24 | 0.36 | 1.18 | 0.43 | 5.07 | 6.88 | 5.87 | 3.04 | 0.85 | 0.68 | 1.62 | 0.94 |
| 14 | 6.3 | 4.4 | 0.38 | 0.33 | 0.34 | 0.34 | 3.68 | 6.17 | 5.86 | 3.37 | 0.80 | 0.94 | 1.30 | 0.76 |
| 15 | 7.7 | 4.5 | 0.48 | 0.21 | 0.57 | 0.32 | 6.31 | 7.45 | 5.77 | 3.22 | 0.77 | 0.57 | 1.55 | 0.76 |
| 16 | 6.5 | 4.9 | 0.26 | 0.23 | 0.36 | 0.35 | 6.19 | 5.86 | 5.95 | 3.21 | 0.15 | 0.75 | 0.96 | 1.06 |
| 17 | 7.7 | 4.6 | 0.29 | 0.24 | 0.88 | 0.35 | 8.77 | 6.52 | 5.67 | 3.37 | 0.77 | 0.74 | 1.55 | 1.02 |
| 18 | 7.9 | 3.8 | 0.37 | 0.34 | 0.68 | 0.40 | 7.94 | 5.96 | 5.71 | 3.28 | 0.71 | 0.73 | 1.63 | 0.75 |
| 19 | 6.7 | 4.5 | 0.22 | 0.29 | 1.02 | 0.31 | 6.33 | 6.32 | 6.79 | 1.12 | 0.55 | 0.73 | 1.02 | 0.82 |
| 20 | 7.2 | 4.5 | 0.44 | 0.31 | 0.32 | 0.38 | 6.96 | 6.46 | 5.57 | 3.36 | 0.68 | 0.70 | 0.92 | 0.57 |
| 21 | 8 | 4.1 | 0.48 | 0.30 | 1.29 | 0.38 | 8.66 | 5.98 | 7.59 | 2.18 | 0.40 | 0.77 | 1.39 | 0.94 |
| 22 | 9 | 3.9 | 0.42 | 0.27 | 0.38 | 0.28 | 4.77 | 6.16 | 3.42 | 3.15 | 1.08 | 0.68 | 1.56 | 0.81 |
| 23 | 8.2 | 4.6 | 0.33 | 0.27 | 0.56 | 0.37 | 4.07 | 5.96 | 2.97 | 3.56 | 1.83 | 0.68 | 1.80 | 1.26 |
| 24 | 8.7 | 3.9 | 0.18 | 0.25 | 1.04 | 0.32 | 5.74 | 6.93 | 2.07 | 3.46 | 0.83 | 0.75 | 2.05 | 0.88 |
| 25 | 8 | 4.3 | 0.21 | 0.29 | 0.98 | 0.42 | 4.28 | 5.78 | 4.92 | 2.99 | 0.70 | 0.60 | 0.45 | 1.18 |
| 26 | 8.6 | 4.4 | 0.31 | 0.20 | 0.89 | 0.34 | 4.58 | 7.48 | 4.92 | 3.66 | 0.77 | 0.71 | 1.42 | 0.45 |

**Supplementary Table 3.**

Key parameters of the vehicle model.

| Parameter | Description | Value |
|---|---|---|
| $m$ | Vehicle mass | 2040 kg |
| $l_f$ | Distances from CoG to front axle | 1.18 m |
| $l_r$ | Distances from CoG to rear axel | 1.72 m |
| $K_f$ | Cornering stiffness of the front axle | $-1.396*10^5$ N/rad |
| $K_r$ | Cornering stiffness of the rear axle | $-1.401*10^5$ N/rad |
| $I_z$ | Vehicle's moment of inertia | 6242 kg·m² |
| $i_{sw}$ | Transmission ratio of the steering system | 16 |
| $I_{sw}$ | The moment of inertia of the steering system | 0.1 kg·m² |
| $B_{sw}$ | Damping coefficient of the steering system | 0.8 N·s/m |
| $K_{sw}$ | Stiffness coefficient of the steering system | 12 N/rad |
| $K_{alig}$ | Gain of self-aligning torque to tire slip angle | -20 N·m/rad |

**Supplementary Table 4.**

Parameters of the MPC controller used to compute $T_{ref}$ in equation (1) in Supplementary Note 2.

| Parameter | Value | Unit |
|---|---|---|
| Weighting factors ($W_1, Q_1$) | ([2.5*$10^3$, 0; 0, 7*$10^3$], 4*$10^2$) | - |
| ($T_{ref,\min}$, $T_{ref,\max}$) | (-10, 10) | N·m |
| ($\Delta T_{ref,\min}$, $\Delta T_{ref,\max}$) | (-10, 10) | N·m/s |
| Prediction horizon $N$ | 10 | - |

**Supplementary Table 5.**

Parameters of the MPC controller used in equation (6).

| Parameter | Value | Unit |
|---|---|---|
| Weighting factors ($W, Q$) | ($10^2$, 1) | - |
| ($T_{hpt,\min}$, $T_{hpt,\max}$) | (-10, 10) | N·m |
| ($\Delta T_{hpt,\min}$, $\Delta T_{hpt,\max}$) | (-10, 10) | N·m/s |
| Time constant $\tau_H$ | 0.5 | - |
| Design gain $\lambda$ | 6 | - |
| Prediction horizon $N$ | 10 | - |